\documentclass[journal]{IEEEtran}
\usepackage{cite}
\usepackage{amsmath,amssymb,amsfonts}
\usepackage{algorithm}  
\usepackage{algpseudocode}  
\usepackage{algpseudocode}
\usepackage{graphicx}
\usepackage{epstopdf}
\usepackage{textcomp}
\usepackage{xcolor}
\usepackage{bm}
\usepackage{setspace}
\allowdisplaybreaks

\ifCLASSINFOpdf
\else
\fi
\begin{document}
	
%
\title{Towards TMA-Based Transmissive RIS Transceiver Enabled Downlink Communication Networks: A Consensus-ADMM Approach}
%
%
%
\author{Zhendong~Li,~Wen~Chen,~\IEEEmembership{Senior~Member,~IEEE},~Haoran~Qin,~Qingqing~Wu,~\IEEEmembership{Senior~Member,~IEEE},~Xusheng~Zhu,~Ziheng~Zhang,~and~Jun~Li,~\IEEEmembership{Senior~Member,~IEEE}
\thanks{This work is supported in part by NSFC 62401448, in part by National key project 2020YFB1807700, NSFC 62071296, and Shanghai Kewei 22JC1404000, in part by NSFC 62371289, NSFC 62331022, and Xiaomi Young Scholar Program, in part by Key Technologies R\&D Program of Jiangsu (Prospective and Key Technologies for Industry) BE2023022, BE2023022-2, and NSFC 62471204. (\emph{Corresponding author: Wen Chen.})}
\thanks{Z. Li is with Department of Electronic Engineering, Shanghai Jiao Tong University, Shanghai 200240 and the School of Information and Communication Engineering, Xi'an Jiaotong University, Xi'an 710049, China (email: lizhendong@xjtu.edu.cn). W. Chen, H. Qin, Q. Wu, X. Zhu, and Z. Zhang are with the Department of Electronic Engineering, Shanghai Jiao Tong University, Shanghai 200240, China (e-mail: wenchen@sjtu.edu.cn; haoranqin@sjtu.edu.cn; qingqingwu@sjtu.edu.cn; xushengzhu@sjtu.edu.cn; zhangziheng@sjtu.edu.cn). J. Li is with the School of Information Science and Engineering, Southeast University, Nanjing 210096, China (Email: jleesr80@gmail.com).}}
\maketitle

\begin{abstract}
	This paper presents a novel multi-stream downlink communication system that utilizes a transmissive reconfigurable intelligent surface (RIS) transceiver. Specifically, we elaborate the downlink communication scheme using time-modulated array (TMA) technology, which enables high order modulation and multi-stream beamforming. Then, an optimization problem is formulated to maximize the minimum signal-to-interference-plus-noise ratio (SINR) with user fairness, which takes into account the constraint of the maximum available power for each transmissive element. Due to the non-convex nature of the formulated problem, finding optimal solution is challenging. To mitigate the complexity, we propose a linear-complexity beamforming algorithm based on consensus alternating direction method of multipliers (ADMM). Specifically, by introducing a set of auxiliary variables, the problem can be decomposed into multiple sub-problems that are amenable to parallel computation, where the each sub-problem can yield closed-form expressions, bringing a significant reduction in the computational complexity. The overall problem achieves convergence by iteratively addressing these sub-problems in an alternating manner. Finally, the convergence of the proposed algorithm and the impact of various parameter configurations on the system performance are validated through numerical simulations.
\end{abstract}


%
\IEEEpeerreviewmaketitle

\section{Introduction}
%
%
%
%
\IEEEPARstart{A}{s} the demand for communication services continues to grow, the next generation of networks beyond the fifth-generation (B5G) and sixth-generation (6G) are anticipated to leverage larger-scale antenna arrays and denser network deployments in order to cater to diverse communication service needs \cite{9530717,9537929}. Nevertheless, these schemes present challenges not only in terms of deployment and implementation but also in terms of their impact on the environment and sustainability. In particular, the deployment of base stations (BSs) equipped with larger-scale antenna arrays necessitates a substantial number of radio frequency (RF) chains and intricate signal processing modules, resulting in a considerable rise in network power consumption \cite{9167258}. Furthermore, the cost of individual BSs is escalating when compared to previous-generation communication networks. Moreover, with B5G and 6G networks anticipated to operate in higher frequency bands and necessitate denser networking, the deployment costs are expected to experience a significant surge. Undoubtedly, power consumption and hardware cost remain prominent issues and challenges in the implementation of future communication networks \cite{9998527,9963962}. Consequently, it becomes imperative to explore innovative network architectures and paradigms that can effectively mitigate power consumption and reduce costs in B5G and 6G networks.

In recent years, the reconfigurable intelligent surface (RIS) has emerged as a revolutionary technology with promising potential to address the challenges encountered in B5G and 6G networks \cite{8811733,9393607,9896755,9485102,9531372,9509394}. The RIS is a planar array comprising a multitude of low-cost passive elements that can be autonomously adjusted by an intelligent controller via control lines. This enables precise regulation of the amplitude and phase shift of incident electromagnetic (EM) waves, empowering the RIS to actively manipulate and enhance wireless communication links. Due to its compact size, the RIS offers a convenient deployment option within networks. By strategically positioning the RIS, it can effectively attenuate interference signals and amplify desired signals, thereby significantly improving the performance of communication systems \cite{9913311}. Furthermore, one notable advantage of the RIS is its inherent almost passive full-duplex operation. In other words, the RIS functions by passively reflecting or transmitting incident signals, eliminating concerns related to self-interference and additional noise commonly associated with conventional relay technology \cite{9174801}. In contrast to conventional multi-antenna systems, the RIS offers a simplified structure that does not necessitate a large number of RF chains. As a result, the power consumption and hardware cost requirements are significantly reduced \cite{9716123}. These inherent advantages have significantly facilitated the deployment and integration of RIS in the upcoming generation of communication networks.

Due to its numerous benefits, RIS has garnered significant attention in the field of wireless communications. RIS technology encompasses three primary modes: reflective RIS, transmissive RIS, and hybrid RIS. These modes have been extensively investigated in various studies, focusing on the enhancement of network performance through joint designs of active and passive beamforming techniques \cite{9365004,9927314,9903905,9133107,9855406,9983541,9200683,9849460,9863732}. Such RIS-assisted communication modes offer promising avenues for improving the overall efficiency and reliability of wireless networks. Apart from its role in auxiliary communication, RIS can also be leveraged as a transceiver for enabling both uplink and downlink communication \cite{9048622,9133266,9570775,9982476}. This distinct approach to RIS utilization expands its potential applications beyond traditional RIS-assisted communication research. Moreover, the exploration of RIS-enabled transceivers is still at an early stage, holding significant promise for cost-effective and energy-efficient communications in B5G and 6G networks. Hence, this factor serves as a strong motivation for the research presented in this paper. 

At present, several studies have proposed the utilization of reflective RIS as a transmitter to enable downlink communication \cite{9133266,7448838}. However, the design of a transmissive RIS transceiver holds greater efficiency compared to a reflective RIS transmitter due to the following primary reasons. Firstly, in reflective RIS transceivers, both the single-antenna horn antenna and the user are situated on the same side of the RIS, whereas in transmissive RIS transceivers, they are positioned on opposite sides of the RIS. As a result, the transmissive RIS transceiver avoids the occurrence of feed source blockage to the incident EM wave, distinguishing it from the reflective RIS transceiver. Additionally, the reflective RIS transceiver introduces echo interference problems due to the incident EM wave and the reflected EM wave being located on the same side. In contrast, the transmissive RIS transceiver effectively avoids this issue by distributing the incident and transmitted waves on different sides of the RIS. This key advantage, along with the elimination of feed source blockage, makes the design of a transmissive RIS transceiver more efficient compared to its reflective transceiver \cite{bai2020high,bai2022time}. Besides, compared to traditional multi-antenna transceivers, the proposed novel transceiver does not require a large number of RF chains and complex signal processing modules. Therefore, its architecture is simpler, and the system design cost and complexity are lower. Hence, it is evident that the design of transmissive RIS transceiver holds promise in enabling efficient, cost-effective, low-power consumption and low-complexity communication in the upcoming B5G and 6G networks.


Building upon the aforementioned background, this paper introduces a novel transmissive RIS transceiver system for downlink communications, focusing on achieving low-cost and low-power consumption. To the best of our knowledge, research in this area is still at an early stage. Furthermore, it is challenging to achieve downlink transmission and beamforming with only a single-antenna horn feed source and passive transmissive RIS. The primary contribution of our proposed scheme lies in leveraging the transmissive RIS to enable an efficient multi-user beamforming scheme using a single active feed horn antenna, which deviates from the conventional multi-antenna transceiver architecture. The key contributions of this paper can be summarized as follows:
\begin{itemize}
    \item We begin by presenting the architecture of the transmissive RIS transceiver enabled downlink communication. We provide a detailed explanation of the mechanism underlying our proposed architecture, highlighting its key distinctions from the conventional transceiver architecture. Due to the passive nature of the transmissive elements in the RIS, they lack the ability for active transmission. To address this, we introduce a scheme that combines high order modulation and multi-stream beamforming to generate control signals for the RIS elements. This scheme is implemented using a time-series array (TMA) technique \cite{9576618}. We further elaborate on the symbol extraction and recovery scheme for signal symbols at the user, ensuring reliable communication.
	
    \item In order to ensure the fairness of users, we formulate an optimization problem to maximize the minimum user signal-to-interference-plus-noise ratio (SINR), thus a low-complexity beamforming scheme for downlink communication systems enabled by the transmissive RIS transceiver is proposed based on consensus alternating direction method of multipliers (ADMM) algorithm \cite{7517329}. More specifically, by introducing a set of auxiliary variables, the problem formulated is decomposed into multiple sub-problems that can be computed in parallel. Since the closed-form expressions of several decomposed sub-problems can be obtained, the computational complexity of the overall problem is significantly reduced to linear complexity. The convergence of the overall problem is achieved by iteratively solving these sub-problems in an alternating manner.
	
    \item Simulation results verify the convergence of the proposed beamforming algorithm based on consensus ADMM, and also illustrate the impact of different parameter configurations based on the proposed algorithm on system performance under this framework. The proposed beamforming algorithm based on consensus ADMM demonstrates favorable convergence properties, while also exhibiting low computational complexity, enabling rapid convergence in more complex network scenarios. Moreover, the TMA-based transmissive RIS transceiver architecture introduced in this study showcases comparable functionality to conventional multi-antenna transceivers. It is noteworthy that the proposed architecture offers advantages in terms of cost-effectiveness and power efficiency. Additionally, scalability is achieved by increasing the number of RIS transmissive elements, facilitating the realization of large-scale antenna systems.
\end{itemize}

The subsequent sections of this paper are structured as follows. In Section II, we provide a comprehensive overview of the downlink communication scheme employing TMA in transmissive RIS transceiver enabled systems. Section III presents the system model and formulates the underlying problem. Building upon this foundation, Section IV presents a detailed exposition of the consensus ADMM-based beamforming algorithm, accompanied by analyses of computational complexity and convergence. In Section V, we present numerical results, which not only validate the convergence and efficacy of our proposed algorithm but also offer valuable insights. Lastly, Section VI concludes this paper by summarizing the key findings and outlining potential avenues for future research.

\textit{Notations:} In this paper, we adhere to a consistent notation convention, where scalars are denoted by lowercase letters, vectors are denoted by bold lowercase letters, and matrices are denoted by bold uppercase letters. The absolute value of a complex-valued scalar $x$ can be denoted by $\left| {x} \right|$, and the Euclidean norm of a complex-valued vector $\bf{x}$ can be denoted by $\left\| {\bf{x}} \right\|$. The conjugates of complex-valued scalar $x$, complex-valued vector $\bf{x}$, and complex-valued matrix $\bf{X}$ are denoted as ${x^ * }$, ${{\bf{x}}^ * }$, and ${{\bf{X}}^ * }$, respectively. In addition, ${\bf{X}}^H$, ${\bf{X}}_{m,n}$ and ${\left\| {\bf{X}} \right\|_F}$ denote rank, conjugate transpose, $m,n$-th entry and Frobenius norm of a matrix $\bf{X}$, respectively. In addition, ${{\mathbf{X}}^{-1}}$ and $\text{vec}(\mathbf{X})$ respectively denote inverse and vectorization of a matrix $\bf{X}$. $\text{diag}(\mathbf{X})$ is a diagonal matrix, and the elements on the diagonal are the matrix $\mathbf{X}$ arranged by column vectorization. ${\mathbb{C}^{M \times N}}$ represents the space of ${M \times N}$ complex matrix. ${\bf{I}}_N$ represents an identity matrix of size ${N \times N}$. $j$ represents the imaginary unit, i.e., $j^2=-1$. Finally, the distribution of a circularly symmetric complex Gaussian (CSCG) random vector with mean $\mu$ and covariance matrix $\bf{C}$ can be expressed as $ {\cal C}{\cal N}\left( {\mu,\bf{C}} \right)$, and $\sim$ denotes `distributed as'. ${\bf{A}} \circ {\bf{B}}$ represents the Hadamard product of the matrices ${\bf{A}}$ and ${\bf{B}}$, and ${\bf{A}} \otimes {\bf{B}}$ represents the Kronecker product of matrices ${\bf{A}}$ and ${\bf{B}}$. 

\section{Downlink Communication Scheme Based TMA for Transmissive RIS Transceiver Enabled System}
We first give the differences and advantages of the proposed architecture compared with other multi-antenna systems. This architecture consists of a single-antenna feed horn antenna and a transmissive RIS equipped with an intelligent controller. Compared to traditional multi-antenna transceiver, it does not require a large number of RF chains and signal processing units, making its architecture simpler, with lower cost and power consumption. Compared to reflective RIS transceiver, it does not suffer from issues such as feed horn blockage and self-interference, resulting in higher radiation efficiency under similar conditions. Next, we present the transmissive RIS transceiver enabled downlink communication systems. Fig. 1 illustrates the architecture, which comprises a feed horn antenna (single antenna) for radiating the carrier, a transmissive RIS equipped with $N$ transmissive phase control elements, and an intelligent controller. Additionally, there are $K$ single-antenna users, with the possibility of extending to accommodate multi-antenna users. The new transceiver configuration comprises both the feed horn antenna and the transmissive RIS. Furthermore, these two components can be integrated into a single device through near-field coupling, thereby reducing the path loss between the feed horn antenna and the transmissive RIS. Note that the downlink employs space division multiple access (SDMA) for multiple users, while the uplink utilizes orthogonal frequency division multiple access (OFDMA) for multiple users. Regarding the multi-stream capability with a single-antenna feed horn antanna, the number of streams it supports is indeed limited due to its processing capabilities. The number of streams supported by the traditional multi-antenna transceiver architecture is related to the number of antennas and their arrangement. Generally, more antennas and more reasonable arrangements can support more independent data streams. In this architecture, different states of each RIS transmissive element correspond to different transmit signals and beamforming designs, so the RIS transmissive elements are equivalent to the antennas of a multi-antenna transceiver. Therefore, by increasing the number of RIS transmissive elements, the number of supported streams can be increased. In addition, increasing the number of feed horn antennas can help to further increase the number of streams.
\begin{figure}
	\centerline{\includegraphics[width=9cm]{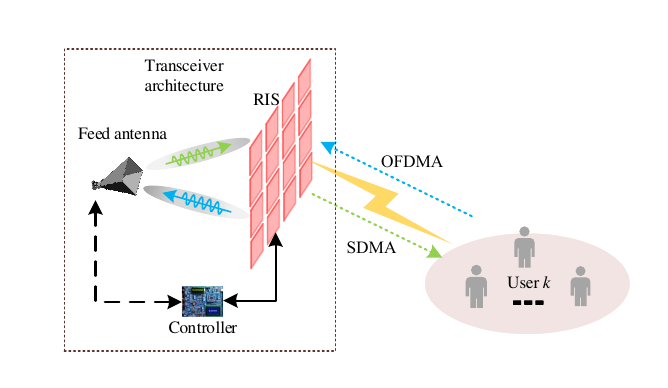}}
	\caption{Transmissive RIS transceiver enabled communication systems.}
	\label{Fig1}
\end{figure}

During the downlink communication process within this architecture, the intelligent controller integrated with the transmissive RIS simultaneously takes into account the multi-stream modulation signals to be transmitted and the beamforming design. It accomplishes this by employing the TMA scheme to generate the appropriate control signals. The TMA scheme enables the intelligent controller to effectively coordinate and optimize the modulation and beamforming aspects, thereby facilitating efficient signal transmission through the transmissive RIS. It is noteworthy that the fundamental principle of the TMA scheme lies in modulating information onto harmonics. Subsequently, the control line facilitates the adjustment of the phase shift of each transmissive element within the RIS. By manipulating these phase shifts, the information is precisely encoded onto the carrier wave emitted by the feed horn antenna. This modulated carrier wave, containing the desired information, is then transmitted to the intended user. Due to the different access methods for the uplink and downlink, this architecture cannot utilize channel reciprocity to acquire channel information. However, by employing a single-antenna feed and a passive transmissive RIS, we have achieved multi-user communication for both uplink and downlink, significantly reducing power consumption and costs. Furthermore, we provide specific channel estimation schemes in the following sections to compensate for the lack of reciprocity in channel acquisition.

Next, we elaborate on the process by which a transmissive RIS transceiver efficiently incorporates multi-stream modulation signals and beamforming to generate the required control signals. Let $s_k$ denote the modulated signal intended for transmission to user $k$, which can be a signal of any modulation order. The composite signal transmitted by the transceiver to all users can be represented as ${\bf{s}} = {\left[ {{s_1}, \ldots, {s_K}} \right]^T} \in {\mathbb{C}^{K \times 1}}$. Moreover, let ${{\bf{f}}_k} = {\left[ {{f_{1,k}}, \ldots, {f_{N,k}}} \right]^T} \in {\mathbb{C}^{N \times 1}}$ represent the beamforming vector for user $k$. The beamforming matrix for all users is denoted as ${\bf{F}} = \left[ {{{\bf{f}}_1}, \ldots, {{\bf{f}}_K}} \right] \in {\mathbb{C}^{N \times K}}$. Hence, the transmitted signal ${\bf{x}} \in {\mathbb{C}^{N \times 1}}$ from the proposed transceiver can be expressed as
\begin{equation}
	{\bf{x}} = {\bf{Fs}}.
\end{equation}
As shown in Eq. (1), the transmission signal $\mathbf{x}$ simultaneously contains the transmitted symbols $\mathbf{s}$ and the beamforming matrix $\mathbf{F}$, forming an $n$-dimensional vector. For any element $x_n$ in this vector, it can be represented as 
\begin{equation}
    \begin{aligned}
        {x_n} &= {\bf{F}}\left( {n,:} \right){\bf{s}} = {a_{n1}}{b_1}{e^{j\left( {{\varphi _{n1}} + {\varpi _1}} \right)}} + ,..., \\
        &+ {a_{nK}}{b_K}{e^{j\left( {{\varphi _{nK}} + {\varpi _K}} \right)}}= {A_n}{e^{j{\phi _n}}},\forall n,
    \end{aligned}
\end{equation}
where $A_n$ and $\phi _n$ represent the signal amplitude and phase of $x_n$, respectively. Then, according to TMA \cite{9576618}, it can generate $n$ control signals for the transmission signal $\mathbf{x}$, corresponding to the control of the $n$ transmissive units of the RIS. Specifically, each element ${\bf{x}}$ of the transmission signal $x_n$ represents the aggregate signal of all users, which can be given by
Taking a 1-bit RIS as an example, the amplitude $A_n$ and phase shift ${\phi _n}$ of the synthesized modulation signal symbol ${x_n}$ can be mapped to control signal waveforms representing two states: 0-state and 1-state, using the TMA scheme. It is important to note that the 0-state and 1-state of the RIS transmissive element correspond to RIS phase shifts of 0 and $\pi$, respectively.

Assume that the symbol time length of the control signal waveform is $T_p$, which is the same and designed uniformly for all elements. Within the symbol time length, the starting moment of the 0-state digital signal relative to the symbol time length is ${t_{on}} $, $0 \le {t_{on}} < {T_p}$. The on-time duration of the 0-state digital signal is $\tau $, $0 \le \tau  < {T_p}$, and all other symbol times are 1-state digital signals. Typical values of the three sets of parameters are as follows: First, $T_p$ is usually related to the performance of the array switch and the number of phase states. The RIS in this paper uses 1 bit, and its value can be about 0.3$\mu s$. Secondly, $t_{on}$ is usually related to the array design, ensuring that the 0-state is entered at the right time to achieve the desired radiation pattern. Its value can be any time point in the range of $\left[0, T_p\right]$. Typical values are some fractional points of $T_p$, such as $T_p/4$, $T_p/2$, etc. Finally, $\tau$ needs to comprehensively consider the sidelobe suppression and radiation pattern optimization, and its typical value can usually be a small part of $T_p$, such as $T_p/10$, $T_p/5$, etc. There are two cases when comparing the sum of the two with the size of $T_p$. If ${t_{on}} + \tau  < {T_p}$, then within a symbol time length $T_p$, after being controlled by digital signal control waveforms in two states of 0 and 1, the phase shift of the transmission electromagnetic (EM) wave can be expressed as
\begin{equation}
    {s_n}(t) = \left\{ \begin{array}{l}
{e^{j0 }},{t_{on,n}} < t \le {t_{on,n}} + {\tau _n},\\
{e^{j\pi }},{\rm{other.}}
\end{array} \right.
\end{equation}
If ${t_{on}} + \tau  > {T_p}$, then within a symbol time length $T_p$, after being controlled by digital signal control waveforms in two states of 0 and 1, the phase shift of the transmission EM wave is the cyclically shifted version of Eq. (3), which can be expressed as
\begin{equation}
     {s_n}(t) = \left\{ \begin{array}{l}
{e^{j0 }},{\rm{other,}}\\
{e^{j\pi }},{t_{on,n}} + {\tau _n} - {T_p} < t \le {t_{on,n}}.
\end{array} \right.
\end{equation}
The control signal waveforms corresponding to the two cases are given in Fig. 2.
\begin{figure}[t]
	\centering 
	\includegraphics[width=8cm]{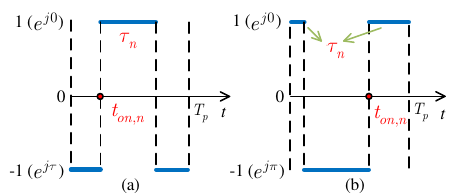}
	\caption{RIS control signal waveforms corresponding to the two cases.}  
	\label{Figure 2} 
\end{figure}Fourier transform the phase control function of the transmission EM wave, and we can obtain
\begin{equation}
    \begin{aligned}
        {S_n}(j2\pi f) = \frac{1}{{{T_p}}}\int_0^{{T_p}} {{s_n}(t){e^{ - j2\pi ft}}dt}  =\\ \left\{ \begin{array}{l}
        \alpha (f,{t_{on,n}},{\tau _n}) + \frac{{j(1 - {e^{ - j2\pi f{T_p}}})}}{{2\pi f{T_p}}},{t_{on,n}} + {\tau _n} \le {T_p},\\
        \alpha (f,{t_{on,n}},{\tau _n} - {T_p}) + \frac{{j({e^{ - j2\pi f{T_p}}} - 1)}}{{2\pi f{T_p}}},{t_{on,n}} + {\tau _n} > {T_p},
        \end{array} \right.
    \end{aligned}
\end{equation}
where $\alpha (f,{t_{on,n}},{\tau _n})$ can be expressed as
\begin{equation}
    \alpha (f,{t_{on,n}},{\tau _n}) = \frac{2}{{\pi f{T_p}}}\sin (\pi f{\tau _n}){e^{ - j\pi f(2{t_{on,n}} + {\tau _n})}}.
\end{equation}
Similar to the Fourier series expansion of time modulation, the peak value of the $q$-th harmonic of the phase modulation function can be expressed as
\begin{equation}
    {\hat S_n}(j2\pi \frac{q}{{{T_p}}}) = \frac{2}{{\pi q}}\sin (\frac{{\pi q{\tau _n}}}{{{T_p}}}){e^{ - j\pi q(2{t_{on,n}} + {\tau _n})/{T_p}}}.
\end{equation}

It should be noted that the amplitude and phase shift of the synthesized modulation signal symbol $x_n$ can be effectively carried by the harmonic component. Moreover, it is important to consider that the positive and negative first harmonic components exhibit the highest energy levels. Therefore, when mapping $x_n$, it is more suitable to allocate it to either the positive first harmonic component or the negative first harmonic component. Consider the positive first harmonic component as an example. Based on the amplitude $A_n$ and phase shift ${\phi _n}$ of the synthesized modulation signal symbol $x_n$, the control signal waveform parameters ${t_{on}}$ and $\tau$ corresponding to the digital signal symbol $x_n$ can be obtained. Specifically, the parameters can be derived as follows:
\begin{equation}
	\frac{{{A_n}}}{{A_n^{\max }}} = \sin \left( {\frac{{\pi \tau }}{{{T_p}}}} \right),\forall n,
\end{equation}
and
\begin{equation}
	- \frac{{\pi \left( {2{t_{on}} + \tau } \right)}}{{{T_p}}} = {\phi _n} + 2k\pi ,0 \le {t_{on}} < {T_p},k \in {\bf{Z}},\forall n,
\end{equation}
where $A_n^{\max}$ represents the maximum amplitude among all symbols in the composite modulation signal symbol $x_n$, and $\mathbf{Z}$ denotes the set of positive integers. In addition, note that the parameters $T_p$, $\tau$, and $t_{on}$ are primarily used to elucidate the generation of downlink control signals based on TMA. As long as these three parameters satisfy the corresponding relationships in the paper, their values do not affect the design of downlink beamforming design below.

Note that the order of harmonic modulation depends on the design of our system. Generally, due to the higher energy of $+1$-order and $-1$-order harmonics and smaller power loss, we generally modulate the signal to be transmitted on $+1$-order or $-1$-order harmonic, and the modulation order used is independent of the number of users. Specifically, we can all use $+1$-order harmonic modulation. After receiving the signal, different users extract the signal on the $+1$-order harmonic. This can be achieved by calculating the harmonic component of the received signal using single-point or two-point fast Fourier transformation (FFT) techniques. Once the composite modulated signal symbols are recovered, the superimposed signals of all users can be obtained. This allows for the application of conventional communication system schemes, such as demodulation and decoding. Hence, the proposed transmissive RIS transceiver architecture can be effectively integrated with existing conventional schemes.

Given that the joint consideration of beamforming and transmission signals in the intelligent controller is integral to generating control signals, the design of the beamforming matrix becomes pivotal in determining the performance of the transmissive RIS transceiver downlink communication systems. In addition, if there is a requirement for a larger number of streams or supporting a greater number of users, we can increase the number of units in the RIS. Then, by generating control signals for RIS units through TMA, it can meet the above requirements. Secondly, if there is a requirement for throughput, in addition to improving performance through beamforming, it can also increase the transmission power of the feed antenna. Therefore, the practical multi-stream capability challenge of this novel architecture can be effectively addressed. Next, we focus on designing the beamforming matrix for the presented communication systems.
\newcounter{my1}
\begin{figure*}[!t]
	\normalsize
	\setcounter{my1}{\value{equation}}
	\setcounter{equation}{15}
        \begin{equation}
	\begin{aligned}
		{{\bf{h}}_{k,{\rm{LoS}}}} &= {{\left[ {1,{e^{ - j\frac{{2\pi }}{\lambda }d\sin \theta _k^{{\rm{AoD}}}\cos \varphi _k^{{\rm{AoD}}}}},...,{e^{ - j\frac{{2\pi }}{\lambda }\left( {{N_x} - 1} \right)d\sin \theta _k^{{\rm{AoD}}}\cos \varphi _k^{{\rm{AoD}}}}}} \right]}^T}\\
		&\otimes {{\left[ {1,{e^{ - j\frac{{2\pi }}{\lambda }d\sin \theta _k^{{\rm{AoD}}}\sin \varphi _k^{{\rm{AoD}}}}},...,{e^{ - j\frac{{2\pi }}{\lambda }\left( {{N_z} - 1} \right)d\sin \theta _k^{{\rm{AoD}}}\sin \varphi _k^{{\rm{AoD}}}}}} \right]}^T},\forall k,
	\end{aligned}
        \end{equation}
	\setcounter{equation}{\value{my1}}
\hrulefill
\vspace*{4pt}
\end{figure*}

\section{System Model and Problem Formulation}
As shown in Fig. 1, this section first introduces the modelling of the proposed transmissive RIS transceiver enabled downlink multi-user (MU) communication system. We represent the channel gain from the transmissive RIS transceiver to the $k$-th user as ${{\bf{h}}_k} \in {\mathbb{C}^{N \times 1}}$ \footnote{Due to the distance between the RIS and the horn feed antenna being less than the Rayleigh distance, we model it as a near-field channel. Meanwhile, the distance between the RIS and the users being greater than the Rayleigh distance, we model it as a far-field channel. During the downlink transmission, the controller jointly performs beamforming and symbols to be transmitted to generate control signals for the RIS units. These control signals are then loaded onto the EM waves emitted by the horn feed antenna and forwarded to the users. Therefore, in the downlink phase, we only need to focus on the far-field channel, as reflected in the modeling part. The near-field channel plays a significant role in the beam alignment process during the uplink phase. For the design of the uplink OFDMA under this architecture, our related work \cite{10242373} has already been published.}. Herein, we give the channel estimation scheme under the novel transmissive RIS-enabled communication system. When the transmissive RIS transceiver transmits the pilot sequence ${{\bf{s}}_k} \in {\mathbb{C}^{L \times 1}}$, the received signal ${{\bf{y}}_k} \in {\mathbb{C}^{L \times 1}}$ at the $k$-th user can be expressed as
\begin{equation}
    {{\bf{y}}_k} = {{\bf{s}}_k}{\bf{f}}_k^H{{\bf{h}}_k} + {{\bf{w}}_k},\forall k,
\end{equation}
with 
\begin{equation}
    {{\bf{w}}_k} = \sum\limits_{i \ne k}^K {{{\bf{s}}_i}{\bf{f}}_i^H{{\bf{h}}_k}}  + {{\bf{n}}_k},\forall k,
\end{equation}
where ${{\bf{w}}_k} \in {\mathbb{C}^{L \times 1}}$ denotes interference of the $k$-th user, and ${{\bf{w}}_k} \sim {\cal C}{\cal N}\left( {{\bf{0}},{{\bf{R}}_k^w}} \right)$. ${\bf{R}}_k^w \in {\mathbb{C}^{L \times L}}$ denotes the covariance matrix of ${{\bf{w}}_k}$, which can be given by
\begin{equation}
    {\bf{R}}_k^w = \sum\limits_{i \ne k}^K {{{\bf{s}}_i}{\bf{f}}_i^H{\bf{R}}_k^h{{\bf{f}}_i}{\bf{s}}_i^H}  + \sigma _k^2{{\bf{I}}_L},\forall k,
\end{equation}
where ${{\bf{R}}_k^h}= \mathbb{E}\left\{ {{{\bf{h}}_k}{\bf{h}}_k^H} \right\} \in {\mathbb{C}^{N \times N}}$. At the $k$-th user, we can apply the minimum mean square error (MMSE) criterion to estimate ${\bf{h}}_k^H$ to obtain ${\bf{\hat h}}_k^H$ as follows
\begin{equation}
    {{{\bf{\hat h}}}_k} = {{\bf{\Xi }}_k}{{\bf{y}}_k},\forall k,
\end{equation}
with
\begin{equation}
    {{\bf{\Xi }}_k} = {\bf{R}}_k^h\left( {{{\bf{f}}_k}{\bf{s}}_k^H} \right){\left[ {\left( {{{\bf{s}}_k}{\bf{f}}_k^H} \right){\bf{R}}_k^h\left( {{{\bf{f}}_k}{\bf{s}}_k^H} \right) + {\bf{R}}_k^w} \right]^{ - 1}},\forall k.
\end{equation}
For the proposed architecture, we first adopt the Lloyd-based codebook design algorithm proposed in \cite{6197256} to synchronize the channel codebook offline at the transmissive RIS transceiver and the users. Then, the above-mentioned channel estimation method is used to obtain the channel state information (CSI) at the user, and then the index information is uploaded when the users request communication to the transmissive RIS transceiver. Note that the final index upload does not rely on channel reciprocity, but is ordinary uplink data transmission. Furthermore, some related works on RIS channel estimation can also provide us with insights \cite{9854847,9521988}. Finally, the beamforming design of the transmissive RIS transceiver is carried out according to the obtained CSI. For the convenience of analysis, it is assumed that all channels exhibit quasi-static flat fading, and that the CSI can be perfectly obtained by the intelligent controller \cite{9982493}. If a more realistic assumption is considered, i.e., the CSI obtained by the system is imperfect, we can use two channel imperfect CSI modeling schemes when modeling, i.e., probability-based and worse case-based. After processing the two constraints, the downlink low-complexity beamforming algorithm framework proposed in this paper can be well applied. A more practical robust beamforming design schemes for imperfect CSI can be considered in future work. The RIS elements are uniformly distributed in a planar array configuration, known as a uniform planar array (UPA). The total number of RIS transmissive elements is denoted as $N = {N_x} \times {N_z}$, where ${N_x}$ and ${N_z}$ represent the number of RIS elements in the horizontal and vertical directions, respectively. Herein, we adopt the Rician channel model to characterize the wireless channel, which can be expressed as
\begin{equation}
	{{\bf{h}}_k} = \sqrt {\beta {{\left( {\frac{{{d_k}}}{{{d_0}}}} \right)}^{ - \alpha }}} \left( {\sqrt {\frac{\kappa }{{\kappa  + 1}}} {{\bf{h}}_{k,{\rm{LoS}}}} + \sqrt {\frac{1}{{\kappa  + 1}}} {{\bf{h}}_{k,{\rm{NLoS}}}}} \right),\forall k,
\end{equation}
where the channel gain at a reference distance ${d_0} = 1{\rm{m}}$ is denoted as $\beta$. The path loss coefficient between the RIS transceiver and the $k$-th user is represented by $\alpha$, while $d_k$ refers to the distance between the RIS transceiver and the $k$-th user. Furthermore, the Rician factor is denoted as $\kappa$. Additionally, the term ${{\bf{h}}_{k,{\rm{LoS}}}}$ represents the line-of-sight (LoS) component of the channel linking the RIS transceiver to the $k$-th user, and its expression is given by Eq. (16), where $\theta _k^{{\rm{AoD}}}$ and $\varphi _k^{{\rm{AoD}}}$ represent the vertical and horizontal angle-of-departure (AoD) at the RIS transceiver, $d$ represents the spacing between the transmissive elements of the RIS, and $\lambda$ corresponds to the wavelength of the carrier. The non-line-of-sight (NLoS) component of the channel between the RIS transceiver and the $k$th user is denoted by ${{\bf{h}}_{k,{\rm{NLoS}}}}$ and ${\left[ {{{\bf{h}}_{k,{\rm{NLoS}}}}} \right]_{\left( {{n_x} - 1} \right){N_z} + {n_z}}} \sim {\cal C}{\cal N}\left( {0,1} \right)$ is the $\left( {{n_x} - 1} \right){N_z} + {n_z}$-th element of ${{\bf{h}}_{k,{\rm{NLoS}}}}$.

Without loss of generality, we make the assumption that the transmission signal is denoted by ${\bf{s}} = {\left[ {{s_1},...,{s_K}} \right]^T} \in {\mathbb{C}^{K \times 1}}$, where ${s_k}\sim {\cal C}{\cal N}\left( {0,1} \right),\forall k$. Hence, the signal received by the $k$-th user can be mathematically expressed as follows
\setcounter{equation}{16}
\begin{equation}
	{y_k} = {\bf{\tilde h}}_k^H\left( {{\rm{vec}}\left( {\bf{F}} \right) \circ {{\bf{a}}_k}} \right) + \sum\limits_{i \ne k}^K {{\bf{\tilde h}}_k^H\left( {{\rm{vec}}\left( {\bf{F}} \right) \circ {{\bf{a}}_i}} \right)}  + {n_k},\forall k,
\end{equation}
where $n_k$ represents the additive white Gaussian noise (AWGN) present at the receiver of the $k$-th user, with ${n_k} \sim {\cal C}{\cal N}\left( {0,\sigma _k^2} \right)$. The vector ${\bf{\tilde h}}_k^H = \left( {{\bf{h}}_k^H,...,{\bf{h}}_k^H} \right) \in {\mathbb{C}^{1 \times NK}}$ denotes the expanded vector obtained from ${\bf{h}}_k^H$. Moreover, ${\rm{vec}}\left( {\bf{F}} \right) = {\left( {{f_{1,1}},...,{f_{N,1}},...,{f_{1,K}},...,{f_{N,K}}} \right)^T} \in {\mathbb{C}^{NK \times 1}}$ represents the vectorized form of the beamforming matrix ${\bf{F}}$. The index vector ${{\bf{a}}_k} \in {\mathbb{C}^{NK \times 1}}$ is specific to the $k$-th user, with a value of 1 at the corresponding index positions within the range $\left( {1,k} \right) \sim \left( {N,k} \right)$ and 0 elsewhere. It can be given by
\begin{equation}
	{{\bf{a}}_k} = {\left( {0,...,0,\underbrace {1,...,1}_{\left( {1,k} \right)\sim \left( {N,k} \right)},0,...,0} \right)^T} \in {\mathbb{R}^{NK \times 1}},\forall k.
\end{equation}
Therefore, the received SINR of the $k$-th user can be mathematically given by
\begin{equation}
	{\rm{SIN}}{{\rm{R}}_k} = \frac{{{{\left| {{\bf{\tilde h}}_k^H\left( {{\rm{vec}}\left( {\bf{F}} \right) \circ {{\bf{a}}_k}} \right)} \right|}^2}}}{{\sum\limits_{i \ne k}^K {{{\left| {{\bf{\tilde h}}_k^H\left( {{\rm{vec}}\left( {\bf{F}} \right) \circ {{\bf{a}}_i}} \right)} \right|}^2} + \sigma _k^2} }},\forall k.
\end{equation}
In addition, it is worth noting that, compared to traditional antennas, the transmissive elements of RIS do not have active transmissive capabilities and only possess passive transmissive capabilities. Since we employ TMA, where information is modulated on harmonics, it is important to note that not all power received by the RIS elements can be effectively utilized. According to the principles of TMA, taking a 1-bit RIS as an example, after modulation, the maximum power that can be applied to each element is $(2/\pi)^2$ the input power, which means a loss of 3.92dB. It is worth noting that in this case, the signal symbol is modulated onto the positive first order harmonic or the negative first order harmonic. If the modulation is on the positive first order harmonic and negative first order harmonic at the same time, the power utilization ratio can be greatly improved \cite{9576618}. Consequently, the transmission power of each element is subject to a constraint imposed by its maximum available power, which corresponds to the power carrying the information. Let us consider the maximum available transmission power for each transmissive element of the RIS, denoted as $P_t$. As a result, the signal transmitted by each transmissive element is subject to the constraint imposed by its maximum available transmission power, which can be given by
\begin{equation}
	{\left\| {{\rm{vec}}\left( {\bf{F}} \right) \circ {{\bf{b}}_n}} \right\|^2} \le {P_t},\forall n,
\end{equation}
where the index vector ${{\bf{b}}_n} \in {\mathbb{C}^{NK \times 1}}$ is associated with the $n$-th RIS element and is defined such that it takes a value of 1 at index positions corresponding to $\left( {n,1} \right),\left( {n,2} \right),...,\left( {n,K} \right)$, and 0 elsewhere. This can be represented as
\begin{equation}
	{{\bf{b}}_n} = {\left( {0,...,\mathop {{\rm{ }}1}\limits_{\left( {n,1} \right)} ,0,...,\mathop {{\rm{ }}1}\limits_{\left( {n,2} \right)} ,0,...,\mathop {{\rm{ }}1}\limits_{\left( {n,K} \right)} ,0,...} \right)^T} \in {\mathbb{R}^{NK \times 1}},\forall n.
\end{equation}

In this paper, we prioritize fairness among all users and present an optimization problem to design the beamforming matrix ${\bf{F}}$ for the transmissive RIS transceiver enabled downlink communication systems\footnote{It is worth noting that the algorithm proposed in this paper can still work for performance metrics considering the overall system performance, such as average rate or system sum-rate. This could be considered as future work for the next stage.}. The objective of the optimization problem, referred to as problem P0, is to maximize the minimum user's SINR. Mathematically, problem P0 can be formulated as 
\begin{subequations}\label{p0}
	\begin{align}
		\text{P0}:\qquad&\mathop {\max }\limits_{\bf{F}} ~{\rm{ min ~ SIN}}{{\rm{R}}_k}, \nonumber\\ 
		\rm{s.t.}\qquad&{\left\| {{\rm{vec}}\left( {\bf{F}} \right) \circ {{\bf{b}}_n}} \right\|^2} \le {P_t},\forall n.
	\end{align}
\end{subequations}

Due to the uncertainty of the convexity or concavity of the objective function with respect to the optimization variable $\bf{F}$, problem P0 is a non-convex optimization problem. Obtaining its optimal solution directly poses challenges. Typically, a common approach is to transform it into a standard semidefinite programming (SDP) problem using techniques such as matrix lifting or SCA. Then, the CVX toolbox is employed to solve it. However, we understand that the computational complexity of this algorithm increases significantly when the number of users and RIS transmissive elements grows. In such cases, obtaining a high-quality solution quickly becomes difficult. In this paper, we propose a consensus ADMM-based approach to solve this problem. The consensus ADMM scheme offers several advantages: firstly, it is designed for distributed optimization, allowing for parallel updates by each set of variables. Then, careful decomposition of the objective function and optimization variables enables efficient updates, potentially even obtaining closed-form solutions, which greatly reduces the computational complexity of the problem. Lastly, as it is based on the general ADMM algorithm framework, its convergence is guaranteed. Next, we provide a detailed beamforming scheme based on consensus ADMM for the transmissive RIS transceiver enabled downlink communications systems.
\section{Consensus ADMM-Based Beamforming for Transmissive RIS Transceiver Enabled Downlink Communication Systems}
\subsection{Problem Transformation}
By introducing the auxiliary variable $\gamma$, problem P0 can be equivalently transformed into problem P1, which is expressed as follows:
\begin{subequations}\label{p1}
	\begin{align}
		\text{P1}:\qquad&\mathop {\max }\limits_{{\bf{F}},\gamma } {\rm{~~}}\gamma , \nonumber\\ 
		\rm{s.t.}\qquad&{\left\| {{\rm{vec}}\left( {\bf{F}} \right) \circ {{\bf{b}}_n}} \right\|^2} \le {P_t},\forall n,\\
		&\frac{{{{\left| {{\bf{\tilde h}}_k^H\left( {{\rm{vec}}\left( {\bf{F}} \right) \circ {{\bf{a}}_k}} \right)} \right|}^2}}}{{\sum\limits_{i \ne k}^K {{{\left| {{\bf{\tilde h}}_k^H\left( {{\rm{vec}}\left( {\bf{F}} \right) \circ {{\bf{a}}_i}} \right)} \right|}^2} + \sigma _k^2} }} \ge \gamma ,\forall k.
	\end{align}
\end{subequations}
It is evident that problem P1 presents a non-convex optimization challenge, making it difficult to obtain the optimal solution directly. To address this issue, this paper proposes the utilization of a low-complexity consensus ADMM algorithm framework, which offers a practical approach to obtain a high-quality solution for this problem.

By applying the consensus ADMM framework and further introducing auxiliary variables $\left\{ {{{\bf{\Gamma }}_n}} \right\},\left\{ {{{\bf{\Psi }}_k}} \right\},\left\{ {{\eta _k}} \right\}$, problem P1 can be equivalently transformed into problem P2, which can be formulated as follows:
\begin{subequations}\label{p2}
	\begin{align}
		\text{P2}:\qquad&\mathop {\max }\limits_{{\bf{F}},\gamma ,\left\{ {{{\bf{\Gamma }}_n}} \right\},\left\{ {{{\bf{\Psi }}_k}} \right\},\left\{ {{\eta _k}} \right\}} {\rm{ }}\gamma , \nonumber\\ 
		\rm{s.t.}\qquad&{\left\| {{\rm{vec}}\left( {{{\bf{\Gamma }}_n}} \right) \circ {{\bf{b}}_n}} \right\|^2} \le {P_t},\forall n,\\
		&\frac{{{{\left| {{\bf{\tilde h}}_k^H\left( {{\rm{vec}}\left( {{{\bf{\Psi }}_k}} \right) \circ {{\bf{a}}_k}} \right)} \right|}^2}}}{{\sum\limits_{i \ne k}^K {{{\left| {{\bf{\tilde h}}_k^H\left( {{\rm{vec}}\left( {{{\bf{\Psi }}_k}} \right) \circ {{\bf{a}}_i}} \right)} \right|}^2} + \sigma _k^2} }} \ge {\eta _k},\forall k,\\
		&{{\bf{\Gamma }}_n} = {\bf{F}},\forall n,\\
		&{{\bf{\Psi }}_k} = {\bf{F}},\forall k,\\
		&{\eta _k} = \gamma ,\forall k.
	\end{align}
\end{subequations}
After auxiliary variables are introduced, problem P2 can be easily decoupled into several sub-problems and obtained the analytical expressions for each sub-problem in parallel.
\subsection{Solve for $\gamma$}
Given ${\bf{F}},\left\{ {{{\bf{\Gamma }}_n}} \right\},\left\{ {{{\bf{\Psi }}_k}} \right\},\left\{ {{\eta _k}} \right\}$, problem P2 can be transformed into problem P3, which can be represented as:
\begin{subequations}\label{p3}
	\begin{align}
		\text{P3}:\qquad&\mathop {\max }\limits_{\gamma } {\rm{~~}}\gamma , \nonumber\\ 
		\rm{s.t.}\qquad&{\eta _k} = \gamma ,\forall k.
	\end{align}
\end{subequations}
Problem P3 can be equivalently transformed into an unconstrained optimization problem P3.1 represented as follows:
\begin{subequations}\label{p3.1}
	\begin{align}
		\text{P3.1}:\qquad&\mathop {\max }\limits_{\gamma ,\left\{ {{\xi _k}} \right\}} {\rm{~~}}\gamma  - \frac{\rho }{2}\sum\limits_{k = 1}^K {{{\left( {{\eta _k} - \gamma  + {\xi _k}} \right)}^2}} ,
	\end{align}
\end{subequations}
where $\rho > 0$ denotes the penalty factor, and $\xi_k$ represents the error term between ${\eta _k}$ and $\gamma$. The objective function of problem P3.1 is defined as $f\left( \gamma \right)$, given by the following expression:
\begin{equation}
	\begin{aligned}
		f\left( \gamma  \right)& = \gamma  - \frac{\rho }{2}\sum\limits_{k = 1}^K {\left( {{\gamma ^2} - 2\left( {{\eta _k} + {\xi _k}} \right)\gamma  + {{\left( {{\eta _k} + {\xi _k}} \right)}^2}} \right)} \\
		& =  - \frac{\rho }{2}K{\gamma ^2} \! + \! \left( {1 \! + \! \rho \sum\limits_{k = 1}^K {\left( {{\eta _k}  \!+ \! {\xi _k}} \right)} } \right)\gamma  \!- \! \frac{\rho }{2}\sum\limits_{k = 1}^K {{{\left( {{\eta _k}  \!+  \!{\xi _k}} \right)}^2}} .
	\end{aligned}
\end{equation}
It can be observed that $f(\gamma)$ is a concave function with respect to (w.r.t) $\gamma$. To determine its maximum value, we differentiate $f(\gamma)$ w.r.t $\gamma$, yielding the expression:
\begin{equation}
	\frac{{df\left( \gamma  \right)}}{{d\gamma }} =  - \rho K\gamma  + \rho \sum\limits_{k = 1}^K {\left( {{\eta _k} + {\xi _k}} \right)}  + 1.
\end{equation}
If we set $\frac{{df\left( \gamma \right)}}{{d\gamma }} = 0$, we can derive the optimal analytical expression for $\gamma$ as follows:
\begin{equation}
	{\gamma  } = \frac{{1 + \rho \sum\limits_{k = 1}^K {\left( {{\eta _k} + {\xi _k}} \right)} }}{{\rho K}},
\end{equation}
where the update criterion of error term ${{\xi _k}}$ will be given later.
\newcounter{my2}
\begin{figure*}[!t]
	\normalsize
	\setcounter{my1}{\value{equation}}
	\setcounter{equation}{42}
        \begin{equation}
	\begin{aligned}
		{{\cal L}_2}\left( {{{\bf{\Psi }}_k}} \right) = \left\| {{{\bf{\Psi }}_k} - {\bf{F}} + {{\bf{\Lambda }}_k}} \right\|_F^2 + {\mu _k}\left( {{\eta _k}\left( {\sum\limits_{i \ne k}^K {{{\left| {{\bf{\tilde h}}_k^H\left( {{\rm{vec}}\left( {{{\bf{\Psi }}_k}} \right) \circ {{\bf{a}}_i}} \right)} \right|}^2} + \sigma _k^2} } \right) - {{\left| {{\bf{\tilde h}}_k^H\left( {{\rm{vec}}\left( {{{\bf{\Psi }}_k}} \right) \circ {{\bf{a}}_k}} \right)} \right|}^2}} \right)\\
	    = {\left\| {{\rm{vec}}\left( {{{\bf{\Psi }}_k} - {\bf{F}} + {{\bf{\Lambda }}_k}} \right)} \right\|^2} + {\mu _k}\left( {{\eta _k}\left( {\sum\limits_{i \ne k}^K {{{\left| {{\bf{\tilde h}}_k^H\left( {{\rm{vec}}\left( {{{\bf{\Psi }}_k}} \right) \circ {{\bf{a}}_i}} \right)} \right|}^2} + \sigma _k^2} } \right) - {{\left| {{\bf{\tilde h}}_k^H\left( {{\rm{vec}}\left( {{{\bf{\Psi }}_k}} \right) \circ {{\bf{a}}_k}} \right)} \right|}^2}} \right),\forall k,
	\end{aligned}
        \end{equation}
        \begin{equation}
	\begin{aligned}
		{\left| {{\bf{\tilde h}}_k^H\left( {{\rm{vec}}\left( {{{\bf{\Psi }}_k}} \right) \circ {{\bf{a}}_k}} \right)} \right|^2}& \ge {\left| {{\bf{\tilde h}}_k^H\left( {{\rm{vec}}\left( {{\bf{\Psi }}_k^{\left( r \right)}} \right) \circ {{\bf{a}}_k}} \right)} \right|^2} + {\left( {\left( {{\bf{H}}_k^T \circ {{\bf{A}}_k}} \right){\rm{vec}}\left( {{{\left( {{\bf{\Psi }}_k^r} \right)}^ * }} \right)} \right)^T}\left( {{\rm{vec}}\left( {{{\bf{\Psi }}_k}} \right) - {\rm{vec}}\left( {{\bf{\Psi }}_k^r} \right)} \right)\\
		& \buildrel \Delta \over = {\left( {{{\left| {{\bf{\tilde h}}_k^H\left( {{\rm{vec}}\left( {{{\bf{\Psi }}_k}} \right) \circ {{\bf{a}}_k}} \right)} \right|}^2}} \right)^{lb}},\forall k.
	\end{aligned}
        \end{equation}
        \begin{equation}
	\begin{aligned}
		{{\cal L}_2}\left( {{{\bf{\Psi }}_k}} \right)& \le {\left\| {{\rm{vec}}\left( {{{\bf{\Psi }}_k} - {\bf{F}} + {{\bf{\Lambda }}_k}} \right)} \right\|^2} + {\mu _k}{\eta _k}\left( {\sum\limits_{i \ne k}^K {{{\left| {{\bf{\tilde h}}_k^H\left( {{\rm{vec}}\left( {{{\bf{\Psi }}_k}} \right) \circ {{\bf{a}}_i}} \right)} \right|}^2} + \sigma _k^2} } \right)\\ &- {\mu _k}{\left( {{{\left| {{\bf{\tilde h}}_k^H\left( {{\rm{vec}}\left( {{{\bf{\Psi }}_k}} \right) \circ {{\bf{a}}_k}} \right)} \right|}^2}} \right)^{lb}} \buildrel \Delta \over = {{\cal L}_2}{\left( {{{\bf{\Psi }}_k}} \right)^{ub}},\forall k.
	\end{aligned}
        \end{equation}
	\setcounter{equation}{\value{my2}}
\hrulefill
\vspace*{4pt}
\end{figure*}
\subsection{Solve for $\left\{ {{{\bf{\Gamma }}_n}} \right\}$}
Given $\gamma ,{\bf{F}},\left\{ {{{\bf{\Psi }}_k}} \right\},\left\{ {{\eta _k}} \right\}$, problem P2 can be transformed into problem P4, which can be expressed as follows:
\setcounter{equation}{29}
\begin{subequations}\label{p4}
	\begin{align}
		\text{P4}:\qquad&\mathop {{\rm{find}}}\limits_{\left\{ {{{\bf{\Gamma }}_n}} \right\}} {\rm{~~}}{{\bf{\Gamma }}_n}, \nonumber\\ 
		\rm{s.t.}\qquad&{\left\| {{\rm{vec}}\left( {{{\bf{\Gamma }}_n}} \right) \circ {{\bf{b}}_n}} \right\|^2} \le {P_t},\forall n,\\
		&{{\bf{\Gamma }}_n} = {\bf{F}},\forall n.
	\end{align}
\end{subequations}
This problem can be equivalently expressed as problem P4.1 as follows:
\begin{subequations}\label{p4.1}
	\begin{align}
		\text{P4.1}:\qquad&\mathop {{\rm{min}}}\limits_{\left\{ {{{\bf{\Gamma }}_n}} \right\},\left\{ {{{\bf{\Xi }}_n}} \right\}} {\rm{ }}\left\| {{{\bf{\Gamma }}_n} - {\bf{F}} + {{\bf{\Xi }}_n}} \right\|_F^2,\nonumber\\ 
		\rm{s.t.}\qquad&{\left\| {{\rm{vec}}\left( {{{\bf{\Gamma }}_n}} \right) \circ {{\bf{b}}_n}} \right\|^2} \le {P_t},\forall n,
	\end{align}
\end{subequations}
where ${{{\bf{\Xi }}_n}}$ represents the error iterm between ${{{\bf{\Gamma }}_n}}$ and $\bf{F}$. The Lagrange function of this problem can be expressed as:
\begin{equation}
	{{\cal L}_1}\left( {{{\bf{\Gamma }}_n}} \right) \!= \!\left\| {{{\bf{\Gamma }}_n} \!-\! {\bf{F}}\! +\! {{\bf{\Xi }}_n}} \right\|_F^2 \!+ \!{\lambda _n}\left( {{{\left\| {{\rm{vec}}\left( {{{\bf{\Gamma }}_n}} \right) \circ {{\bf{b}}_n}} \right\|}^2}\! -\! {P_t}} \right),\forall n,
\end{equation}
where ${\lambda _n}$ is the Lagrangian multiplier associated with the constraint (31a). It can be observed that ${{\cal L}_1}\left( {{{\bf{\Gamma }}_n}} \right)$ is convex w.r.t ${{{\bf{\Gamma }}_n}}$. We take the partial derivative of ${{\cal L}_1}\left( {{{\bf{\Gamma }}_n}} \right)$ w.r.t ${{{\bf{\Gamma }}_n}}$, denoted as
\begin{equation}
	\frac{{\partial {{\cal L}_1}\left( {{{\bf{\Gamma }}_n}} \right)}}{{\partial {{\bf{\Gamma }}_n}}} = {\left( {{{\bf{\Gamma }}_n} - {\bf{F}} + {{\bf{\Xi }}_n}} \right)^ * } + {\lambda _n}{\bf{\Gamma }}_{_n}^ *  \circ {{\bf{B}}_n},\forall n,
\end{equation}
where ${{\bf{B}}_n} \in {\mathbb{R}^{N \times K}}$ is a matrix with 1 in the $n$-th row and 0 elsewhere.

\emph{Proof:} See Appendix A. $\hfill\blacksquare$ 

Let $\frac{{\partial {{\cal L}_1}\left( {{{\bf{\Gamma }}_n}} \right)}}{{\partial {{\bf{\Gamma }}_n}}} = 0$, i.e.,
\begin{equation}
	{\bf{\Gamma }}_n^ *  - {{\bf{F}}^ * } + {\bf{\Xi }}_n^ *  + {\lambda _n}{\bf{\Gamma }}_{_n}^ *  \circ {{\bf{B}}_n} = 0,\forall n.
\end{equation}
Taking the conjugate of both sides of Eq. (34), we can obtain
\begin{equation}
	{{\bf{\Gamma }}_n} - {\bf{F}} + {{\bf{\Xi }}_n} + {\lambda _n}{{\bf{\Gamma }}_n} \circ {{\bf{{ B}}}_n} = 0,\forall n.
\end{equation}
The expression can be further simplified as follows:
\begin{equation}
	\left( {{\bf{1}} + {\lambda _n}{{\bf{B}}_n}} \right) \circ {{\bf{\Gamma }}_n} = {\bf{F}} - {{\bf{\Xi }}_n},\forall n,
\end{equation}
where the elements of the matrix ${\bf{1}} \in {\mathbb{R}^{N \times K}}$ are all one. Then, in order to obtain ${{\bf{\Gamma }}_n}$, we vectorize both sides of the Eq. (36), yielding
\begin{equation}
	{\rm{vec}}\left( {\left( {{\bf{1}} + {\lambda _n}{{\bf{B}}_n}} \right) \circ {{\bf{\Gamma }}_n}} \right) = {\rm{vec}}\left( {{\bf{F}} - {{\bf{\Xi }}_n}} \right),\forall n,
\end{equation}

{\bf{Theorem 1:}} \emph{Given matrices ${\bf{A}} \in {\mathbb{C}^{N \times K}}$ and ${\bf{X}} \in {\mathbb{C}^{N \times K}}$,  it can be shown that ${\rm{vec}}\left( {{\bf{A}} \circ {\bf{X}}} \right) = {\rm{diag}}\left( {\bf{A}} \right){\rm{vec}}\left( {\bf{X}} \right)$ holds.}

According to {\bf{Theorem 1}}, Eq. (37) can be further expressed as
\begin{equation}
	{\rm{diag}}\left( {{\bf{1}} + {\lambda _n}{{\bf{B}}_n}} \right){\rm{vec}}\left( {{{\bf{\Gamma }}_n}} \right) = {\rm{vec}}\left( {{\bf{F}} - {{\bf{\Xi }}_n}} \right),\forall n.
\end{equation}
Thus, the optimal analytical expression of ${{{\bf{\Gamma }}_n}}$ can be denoted by
\begin{equation}
	{\rm{vec}}\left( {{\bf{\Gamma }}_n} \right) = {\left( {{\rm{diag}}\left( {{\bf{1}} + {\lambda _n}{{\bf{B}}_n}} \right)} \right)^{ - 1}}{\rm{vec}}\left( {{\bf{F}} - {{\bf{\Xi }}_n}} \right),\forall n,
\end{equation}
where the update of ${{{\bf{\Xi }}_n}}$ will be given later, and the update of Lagrangian multiplier ${\lambda _n}$ adopts the following criteria:
\begin{equation}
	\lambda _n^i = {\left[ {\lambda _n^{i - 1} - {\alpha _n}\left( {{{\left\| {{\rm{vec}}\left( {{{\bf{\Gamma }}_n}} \right) \circ {{\bf{b}}_n}} \right\|}^2} - {P_t}} \right)} \right]^ + }, \forall n,
\end{equation}
where ${\alpha _n}$ denotes the step size, and ${\left[  \cdot  \right]^ + } = \max \left\{ {0, \cdot } \right\}$. 

\newcounter{my3}
\begin{figure*}[!t]
	\normalsize
	\setcounter{my3}{\value{equation}}
	\setcounter{equation}{47}
        \begin{equation}
        \begin{array}{l}
		  \mu _k^i = {\left[ {\mu _k^{i - 1} - {\beta _k}\left( {{\eta _k}\left( {\sum\limits_{i \ne k}^K {{{\left| {{\bf{\tilde h}}_k^H\left( {{\rm{vec}}\left( {{\bf{\Psi }}_k^ * } \right) \circ {{\bf{a}}_i}} \right)} \right|}^2} + \sigma _k^2} } \right) - {{\left( {{{\left| {{\bf{\tilde h}}_k^H\left( {{\rm{vec}}\left( {{\bf{\Psi }}_k^ * } \right) \circ {{\bf{a}}_k}} \right)} \right|}^2}} \right)}^{lb}}} \right)} \right]^ + },\forall k,
	\end{array}
        \end{equation}
        \setcounter{equation}{50}
        \begin{equation}
        {{\cal L}_3}\left( {{\eta _k}} \right) = {\left( {{\eta _k} - \gamma  + {\xi _k}} \right)^2} + {\theta _k}\left( {{\eta _k}\left( {\sum\limits_{i \ne k}^K {{{\left| {{\bf{\tilde h}}_k^H\left( {{\rm{vec}}\left( {{{\bf{\Psi }}_k}} \right) \circ {{\bf{a}}_i}} \right)} \right|}^2} + \sigma _k^2} } \right) - {{\left| {{\bf{\tilde h}}_k^H\left( {{\rm{vec}}\left( {{{\bf{\Psi }}_k}} \right) \circ {{\bf{a}}_k}} \right)} \right|}^2}} \right), \forall k,    
        \end{equation}
        \begin{equation}
        \frac{{d{{\cal L}_3}\left( {{\eta _k}} \right)}}{{d{\eta _k}}} = 2\left( {{\eta _k} - \gamma  + {\xi _k}} \right) + {\theta _k}\left( {\sum\limits_{i \ne k}^K {{{\left| {{\bf{\tilde h}}_k^H\left( {{\rm{vec}}\left( {{{\bf{\Psi }}_k}} \right) \circ {{\bf{a}}_i}} \right)} \right|}^2} + \sigma _k^2} } \right),\forall k.
        \end{equation}
        \setcounter{equation}{53}
        \begin{equation}
        \theta _k^i = {\left[ {\theta _k^{i - 1} - {\tau _k}\left( {{\eta _k}\left( {\sum\limits_{i \ne k}^K {{{\left| {{\bf{\tilde h}}_k^H\left( {{\rm{vec}}\left( {{\bf{\Psi }}_k^ * } \right) \circ {{\bf{a}}_i}} \right)} \right|}^2} + \sigma _k^2} } \right) - {{\left| {{\bf{\tilde h}}_k^H\left( {{\rm{vec}}\left( {{\bf{\Psi }}_k^ * } \right) \circ {{\bf{a}}_k}} \right)} \right|}^2}} \right)} \right]^ + },\forall k,
\end{equation}
\setcounter{equation}{\value{my3}}
\hrulefill
\vspace*{4pt}
\end{figure*}
\subsection{Solve for $\left\{ {{{\bf{\Psi }}_k}} \right\}$}
Given $\gamma ,{\bf{F}},\left\{ {{{\bf{\Gamma }}_n}} \right\},\left\{ {{\eta _k}} \right\}$, problem P2 can be transformed into problem P5, which can be represented as follows:
\begin{subequations}\label{p5}
	\begin{align}
		\text{P5}:\qquad&\mathop {{\rm{find}}}\limits_{\left\{ {{{\bf{\Psi }}_k}} \right\}} {\rm{~~}}{{\bf{\Psi }}_k}, \nonumber\\ 
		\rm{s.t.}\qquad&{\left| {{\bf{\tilde h}}_k^H\left( {{\rm{vec}}\left( {{{\bf{\Psi }}_k}} \right) \circ {{\bf{a}}_k}} \right)} \right|^2} \ge \nonumber\\
        &{\eta _k}\left( {\sum\limits_{i \ne k}^K {{{\left| {{\bf{\tilde h}}_k^H\left( {{\rm{vec}}\left( {{{\bf{\Psi }}_k}} \right) \circ {{\bf{a}}_i}} \right)} \right|}^2} + \sigma _k^2} } \right),\forall k,\\
		&{{\bf{\Psi }}_k} = {\bf{F}},\forall k.
	\end{align}
\end{subequations}
This problem can be equivalently transformed into problem P5.1, which can be denoted by
\begin{subequations}\label{p5.1}
	\begin{align}
		\text{P5.1}:\qquad&\mathop {{\rm{min}}}\limits_{\left\{ {{{\bf{\Psi }}_k}} \right\},\left\{{{\bf{\Lambda }}_k}\right\}} {\rm{ }}\left\| {{{\bf{\Psi }}_k} - {\bf{F}} + {{\bf{\Lambda }}_k}} \right\|_F^2, \nonumber\\ 
		\rm{s.t.}\qquad&{\left| {{\bf{\tilde h}}_k^H\left( {{\rm{vec}}\left( {{{\bf{\Psi }}_k}} \right) \circ {{\bf{a}}_k}} \right)} \right|^2} \ge \nonumber\\
        &{\eta _k}\left( {\sum\limits_{i \ne k}^K {{{\left| {{\bf{\tilde h}}_k^H\left( {{\rm{vec}}\left( {{{\bf{\Psi }}_k}} \right) \circ {{\bf{a}}_i}} \right)} \right|}^2} + \sigma _k^2} } \right),\forall k,
	\end{align}
\end{subequations}
where ${{{\bf{\Lambda }}_k}}$ represents the error term between ${{{\bf{\Psi }}_k}}$ and $\bf{F}$. The Lagrange function of this problem can be expressed as Eq. (43), where ${\mu _k}$ is the Lagrangian multiplier associated with the constraint (42a). It can be observed that ${{\cal L}_2}\left( {{{\bf{\Psi }}_k}} \right)$ is non-convex w.r.t ${\rm{vec}}\left( {{{\bf{\Psi }}_k}} \right)$. At the $r$-th iteration, we can use SCA to obtain a lower bound for ${\left| {{\bf{\tilde h}}_k^H\left( {{\rm{vec}}\left( {{{\bf{\Psi }}_k}} \right) \circ {{\bf{a}}_k}} \right)} \right|^2}$, which can be expressed as Eq. (44).

\emph{Proof:} See Appendix B. $\hfill\blacksquare$ 

Therefore, we have Eq. (45). It can be observed that ${\cal L}{\left( {{{\bf{\Psi }}_k}} \right)^{ub}}$ is convex w.r.t ${{\rm{vec}}\left( {{{\bf{\Psi }}_k}} \right)}$. In order to obtain its minimum value, the partial derivative of ${\cal L}{\left( {{{\bf{\Psi }}_k}} \right)^{ub}}$ w.r.t ${{\rm{vec}}\left( {{{\bf{\Psi }}_k}} \right)}$ can be obtained as follows:
\setcounter{equation}{45}
\begin{equation}
    \begin{aligned}
        \frac{{\partial {{\cal L}_2}{{\left( {{{\bf{\Psi }}_k}} \right)}^{ub}}}}{{\partial {\rm{vec}}\left( {{{\bf{\Psi }}_k}} \right)}} &= {\rm{vec}}\left( {{\bf{\Psi }}_k^ *  - {{\bf{F}}^ * } + {\bf{\Lambda }}_k^ * } \right)\\
        &+ {\mu _k}{\eta _k}\sum\limits_{i \ne k}^K {\left( {{\bf{H}}_k^T \circ {{\bf{A}}_i}} \right){\rm{vec}}\left( {{\bf{\Psi }}_k^ * } \right)}  \\
        &- {\mu _k}\left( {{\bf{H}}_k^T \circ {{\bf{A}}_k}} \right){\rm{vec}}\left( {{{\left( {{\bf{\Psi }}_k^r} \right)}^ * }} \right),\forall k.
    \end{aligned}
\end{equation}

\emph{Proof:} See Appendix B. $\hfill\blacksquare$ 

Let $\frac{{\partial {{\cal L}_2}{{\left( {{{\bf{\Psi }}_k}} \right)}^{ub}}}}{{\partial {\rm{vec}}\left( {{{\bf{\Psi }}_k}} \right)}} = 0$, we can obtain the optimal analytical expression of ${{{\bf{\Psi }}_k}}$ as follows:
\begin{equation}
    \begin{aligned}
	{\rm{vec}}\left( {{\bf{\Psi }}_k} \right) = {\left( {{{\bf{I}}_{NK}} + {\mu _k}{\eta _k}\sum\limits_{i \ne k}^K {\left( {{{\bf{H}}_k} \circ {{\bf{A}}_i}} \right)} } \right)^{ - 1}} \cdot \\\left( {{\mu _k}\left( {{{\bf{H}}_k} \circ {{\bf{A}}_k}} \right){\rm{vec}}\left( {{\bf{\Psi }}_k^r} \right) + {\rm{vec}}\left( {{\bf{F}} - {{\bf{\Lambda }}_k}} \right)} \right),\forall k,
    \end{aligned}
\end{equation}
where the update criteria of ${{{\bf{\Lambda }}_k}}$ are given later, and the update of Lagrangian multiplier ${\mu _k}$ is represented as Eq. (48), where ${\beta _k}$ denotes the step size.

\subsection{Solve for $\left\{ {{\eta _k}} \right\}$}
Given $\gamma ,{\bf{F}},\left\{ {{{\bf{\Gamma }}_n}} \right\},\left\{ {{{\bf{\Psi }}_k}} \right\}$, problem P2 can be rewritten as problem P6 as follows:
\setcounter{equation}{48}
\begin{subequations}\label{p6}
	\begin{align}
		\text{P6}:\qquad&\mathop {{\rm{find}}}\limits_{\left\{ {{\eta _k}} \right\}} {\rm{~~}}{\eta _k}, \nonumber\\ 
		\rm{s.t.}\qquad&\frac{{{{\left| {{\bf{\tilde h}}_k^H\left( {{\rm{vec}}\left( {{{\bf{\Psi }}_k}} \right) \circ {{\bf{a}}_k}} \right)} \right|}^2}}}{{\sum\limits_{i \ne k}^K {{{\left| {{\bf{\tilde h}}_k^H\left( {{\rm{vec}}\left( {{{\bf{\Psi }}_k}} \right) \circ {{\bf{a}}_i}} \right)} \right|}^2} + \sigma _k^2} }} \ge {\eta _k},\forall k,\\
		&{\eta _k} = \gamma ,\forall k.
	\end{align}
\end{subequations}
Further, problem P6 can be equivalently transformed into problem P6.1 as follows
\begin{subequations}\label{p6.1}
	\begin{align}
		\text{P6.1}:\qquad&\mathop {\min }\limits_{\left\{ {{\eta _k}} \right\}, \left\{ {{\xi _k}} \right\}} {\rm{~~}}{\left( {{\eta _k} - \gamma  + {\xi _k}} \right)^2}, \nonumber\\ 
		\rm{s.t.}\qquad&\frac{{{{\left| {{\bf{\tilde h}}_k^H\left( {{\rm{vec}}\left( {{{\bf{\Psi }}_k}} \right) \circ {{\bf{a}}_k}} \right)} \right|}^2}}}{{\sum\limits_{i \ne k}^K {{{\left| {{\bf{\tilde h}}_k^H\left( {{\rm{vec}}\left( {{{\bf{\Psi }}_k}} \right) \circ {{\bf{a}}_i}} \right)} \right|}^2} + \sigma _k^2} }} \ge {\eta _k},\forall k,
	\end{align}
\end{subequations}
where $\xi_k$ represents the error term between ${\eta _k}$ and $\gamma$. The Lagrange function corresponding to this problem can be expressed as Eq. (51), where ${\theta _k}$ is the Lagrangian multiplier associated with the constraint (50a). Since ${{\cal L}_3}\left( {{\eta _k}} \right)$ is a convex function w.r.t ${{\eta _k}}$, in order to obtain its minimum value, we differentiate ${{\cal L}_3}\left( {{\eta _k}} \right)$ to ${{\eta _k}}$, which can be expressed as Eq. (52). Let $\frac{{d {{\cal L}_3}\left( {{\eta _k}} \right)}}{{d {\eta _k}}} = 0$, the optimal analytical expression of ${{\eta _k}}$ can be obtained as follows:
\setcounter{equation}{52}
\begin{equation}
	\eta _k  = \gamma  - {\xi _k} - \frac{{{\theta _k}}}{2}\left( {\sum\limits_{i \ne k}^K {{{\left| {{\bf{\tilde h}}_k^H\left( {{\rm{vec}}\left( {{{\bf{\Psi }}_k}} \right) \circ {{\bf{a}}_i}} \right)} \right|}^2} + \sigma _k^2} } \right),\forall k,
\end{equation}
where the update scheme of ${\xi _k}$ will be given later, and the update of Lagrangian multiplier ${\theta _k}$ is represented as Eq. (54), where ${\tau _k}$ denotes the step size.
\subsection{Solve for $\bf{F}$}
Given $\gamma ,\left\{ {{{\bf{\Gamma }}_n}} \right\},\left\{ {{{\bf{\Psi }}_k}} \right\},{\eta _k}$, problem P2 can be transformed into problem P7, which can be represented as follows:
\setcounter{equation}{54}
\begin{subequations}\label{p7}
	\begin{align}
		\text{P7}:\qquad&\mathop {{\rm{find}}}\limits_{\bf{F}} {\rm{~~}}{\bf{F}}, \nonumber\\ 
		\rm{s.t.}\qquad&{{\bf{\Gamma }}_n} = {\bf{F}},\forall n,\\
		&{{\bf{\Psi }}_k} = {\bf{F}},\forall k.
	\end{align}
\end{subequations}
Problem P7 can be further converted into unconstrained problem P7.1, expressed as follows:
\begin{subequations}\label{p7.1}
	\begin{align}
		\text{P7.1}:\qquad&\mathop {\min }\limits_{{\bf{F}},\left\{ {{{\bf{\Lambda }}_k}} \right\},\left\{ {{{\bf{\Xi }}_n}} \right\}} {\rm{ }}\sum\limits_{k = 1}^K {\left\| {{{\bf{\Psi }}_k} - {\bf{F}} + {{\bf{\Lambda }}_k}} \right\|_F^2}  + \nonumber\\
        &\sum\limits_{n = 1}^N {\left\| {{{\bf{\Gamma }}_n} - {\bf{F}} + {{\bf{\Xi }}_n}} \right\|_F^2} ,
	\end{align}
\end{subequations}
where ${{{\bf{\Xi }}_n}}$ represents the error iterm between ${{{\bf{\Gamma }}_n}}$ and $\bf{F}$, and ${{{\bf{\Lambda }}_k}}$ denotes the error iterm between ${{{\bf{\Psi }}_k}}$ and $\bf{F}$. We define that
\begin{equation}
	h\left( {\bf{F}} \right) = \sum\limits_{k = 1}^K {\left\| {{{\bf{\Psi }}_k} - {\bf{F}} + {{\bf{\Lambda }}_k}} \right\|_F^2}  + \sum\limits_{n = 1}^N {\left\| {{{\bf{\Gamma }}_n} - {\bf{F}} + {{\bf{\Xi }}_n}} \right\|_F^2} .
\end{equation}
Since $h\left( {\bf{F}} \right)$ is convex w.r.t $\bf{F}$, in order to obtain its minimum value, we take the partial derivative of $\bf{F}$, which can be expressed as follows:
\begin{equation}
	\frac{{\partial h\left( {\bf{F}} \right)}}{{\partial {\bf{F}}}} = \sum\limits_{k = 1}^K {{{\left( {{{\bf{\Psi }}_k} - {\bf{F}} + {{\bf{\Lambda }}_k}} \right)}^ * }}  + \sum\limits_{n = 1}^N {{{\left( {{{\bf{\Gamma }}_n} - {\bf{F}} + {{\bf{\Xi }}_n}} \right)}^ * }} .
\end{equation}
Let $\frac{{\partial h\left( {\bf{F}} \right)}}{{\partial {\bf{F}}}} = 0$, we have
\begin{equation}
	\sum\limits_{k = 1}^K {{{\left( {{{\bf{\Psi }}_k} - {\bf{F}} + {{\bf{\Lambda }}_k}} \right)}^ * }}  + \sum\limits_{n = 1}^N {{{\left( {{{\bf{\Gamma }}_n} - {\bf{F}} + {{\bf{\Xi }}_n}} \right)}^ * }}  = 0,
\end{equation}
which can be further expressed as
\begin{equation}
	\left( {K + N} \right){\bf{F}} = \sum\limits_{k = 1}^K {\left( {{{\bf{\Psi }}_k} + {{\bf{\Lambda }}_k}} \right)}  + \sum\limits_{n = 1}^N {\left( {{{\bf{\Gamma }}_n} + {{\bf{\Xi }}_n}} \right)} .
\end{equation}
Therefore, the optimal analytical expression of $\bf{F}$ can be obtained as follows:
\begin{equation}
	{{\bf{F}}} = \frac{1}{{K + N}}\left( {\sum\limits_{k = 1}^K {\left( {{{\bf{\Psi }}_k} + {{\bf{\Lambda }}_k}} \right)}  + \sum\limits_{n = 1}^N {\left( {{{\bf{\Gamma }}_n} + {{\bf{\Xi }}_n}} \right)} } \right),
\end{equation}
where the update scheme of ${{{\bf{\Lambda }}_k}}$ and ${{{\bf{\Xi }}_n}}$ will be given later.
\subsection{Update of Error Term}
The updates for the error terms $\xi_k$, ${{{\bf{\Lambda }}_k}}$, and ${{{\bf{\Xi }}_n}}$ mentioned earlier are as follows:
\begin{equation}
	{\xi _k}\leftarrow {\xi _k} + {\eta _k} - \gamma ,\forall k,
\end{equation}
\begin{equation}
	{{\bf{\Xi }}_n} \leftarrow {{\bf{\Xi }}_n} + {{\bf{\Gamma }}_n} - {\bf{F}},\forall n,
\end{equation}
\begin{equation}
	{{\bf{\Lambda }}_k} \leftarrow {{\bf{\Lambda }}_k} + {{\bf{\Psi }}_k} - {\bf{F}},\forall k.
\end{equation}
\subsection{Overall Consensus ADMM-Based Beamforming Design Algorithm}
Based on the preceding sub-problems, we present {\bf{Algorithm 1}}, which outlines the proposed comprehensive approach for the beamforming design in transmissive RIS transceiver enabled downlink communication systems. Specifically, the optimization variables are determined iteratively by solving each sub-problem independently, while keeping the values of other variables fixed. It is important to ensure the convergence of the Lagrangian multipliers in the Lagrangian problem. The error term is updated accordingly, and the process continues by iteratively solving the sub-problems until global convergence of the overall problem is achieved. For a more detailed description, please refer to {\bf{Algorithm 1}}.
\begin{algorithm}[]
	\caption{Consensus ADMM-Based Beamforming Algorithm} 
	\begin{algorithmic}[1]
		\State$\textbf{Input:}$ ${\gamma ^0}$, ${{\bf{\Gamma }}_n^0}$, ${\bf{\Psi }}_k^0$, $\eta _k^0$, ${{\bf{F}}^0}$, $\xi _k^0$, ${\bf{\Xi }}_n^0$, ${\bf{\Lambda }}_k^0$, $\lambda _n^0$, $\mu _k^0$, $\theta _k^0$, $\rho $, ${\alpha _n}$, ${\beta _k}$, $\tau _k$, threshold $\epsilon$ and iteration index $r = 0$.
		\Repeat
		\State Obtain ${\gamma  }$ according to Eq. (29).
		\State Obtain ${{\bf{\Gamma }}_n}$ according to Eq. (39) and Eq. (40).
		\State Obtain ${{\bf{\Psi }}_k}$ according to Eq. (47) and Eq. (48).
		\State Obtain ${\eta _k}$ according to Eq. (53) and Eq. (54).
		\State Obtain ${\bf{F }}$ according to Eq. (61).
		\State Update error term $\xi_k$, ${{{\bf{\Lambda }}_k}}$, and ${{{\bf{\Xi }}_n}}$ according to Eq. (62), Eq. (63) and Eq. (64), respectively.
		\State Update iteration index $r=r+1$.
		\Until The fractional decrease of the objective value is below a threshold $\epsilon$.
		\State \Return Beamforming scheme.
	\end{algorithmic}
\end{algorithm}

\subsection{Convergence and Computational Complexity Analysis}
\subsubsection{Convergence Analysis}
In this paper, our primary focus is on analyzing the convergence of the proposed algorithm applied to the transformed problem P2. We define the objective function of problem P2 as $\gamma \left( {{\bf{F}},\gamma ,\left\{ {{{\bf{\Gamma }}_n}} \right\},\left\{ {{{\bf{\Psi }}_k}} \right\},\left\{ {{\eta _k}} \right\}} \right)$. Furthermore, we use ${\gamma ^r}$, ${\left\{ {{{\bf{\Gamma }}_n}} \right\}^r}$, ${\left\{ {{{\bf{\Psi }}_k}} \right\}^r}$, ${\left\{ {{\eta _k}} \right\}^r}$, and ${{\bf{F}}^r}$ to represent the solutions of each subproblem at the $r$-th iteration. In the third step of {\bf{Algorithm 1}}, given ${\left\{ {{{\bf{\Gamma }}_n}} \right\}^r}$, ${\left\{ {{{\bf{\Psi }}_k}} \right\}^r}$, ${\left\{ {{\eta _k}} \right\}^r}$, and ${\bf{F}}^r$, the value of variable ${\gamma ^{r+1}}$ can be obtained, therefore, the following formula holds:
\begin{equation}
\begin{aligned}
    \gamma \left( {{{\bf{F}}^r},{\gamma ^r},{{\left\{ {{{\bf{\Gamma }}_n}} \right\}}^r},{{\left\{ {{{\bf{\Psi }}_k}} \right\}}^r},{{\left\{ {{\eta _k}} \right\}}^r}} \right) \le\\ \gamma \left( {{{\bf{F}}^r},{\gamma ^{r + 1}},{{\left\{ {{{\bf{\Gamma }}_n}} \right\}}^r},{{\left\{ {{{\bf{\Psi }}_k}} \right\}}^r},{{\left\{ {{\eta _k}} \right\}}^r}} \right).
\end{aligned}
\end{equation}
Then, in the fourth step to the seventh step of {\bf{Algorithm 1}}, the four sub-problems it solves are all a feasibility-check problem, so the value of the objective function will not change. Overall, the following formula holds:
\begin{equation}
\begin{aligned}
    \gamma \left( {{{\bf{F}}^r},{\gamma ^{r + 1}},{{\left\{ {{{\bf{\Gamma }}_n}} \right\}}^r},{{\left\{ {{{\bf{\Psi }}_k}} \right\}}^r},{{\left\{ {{\eta _k}} \right\}}^r}} \right) \le\\ \gamma \left( {{{\bf{F}}^{r + 1}},{\gamma ^{r + 1}},{{\left\{ {{{\bf{\Gamma }}_n}} \right\}}^{r + 1}},{{\left\{ {{{\bf{\Psi }}_k}} \right\}}^{r + 1}},{{\left\{ {{\eta _k}} \right\}}^{r + 1}}} \right).
\end{aligned}
\end{equation}
Hence, we have
\begin{equation}
\begin{aligned}
    \gamma \left( {{{\bf{F}}^r},{\gamma ^r},{{\left\{ {{{\bf{\Gamma }}_n}} \right\}}^r},{{\left\{ {{{\bf{\Psi }}_k}} \right\}}^r},{{\left\{ {{\eta _k}} \right\}}^r}} \right) \le \\\gamma \left( {{{\bf{F}}^{r + 1}},{\gamma ^{r + 1}},{{\left\{ {{{\bf{\Gamma }}_n}} \right\}}^{r + 1}},{{\left\{ {{{\bf{\Psi }}_k}} \right\}}^{r + 1}},{{\left\{ {{\eta _k}} \right\}}^{r + 1}}} \right),
\end{aligned}
\end{equation}
which demonstrates that the objective function value in each iteration of the {\bf{Algorithm 1}} is non-decreasing. Furthermore, since the objective function of problem (P2) has an upper bound, the convergence of {\bf{Algorithm 1}} can be assured.
\subsubsection{Computational complexity analysis}
The computational complexity of {\bf{Algorithm 1}} primarily relies on solving sub-problems 4.1, 5.1, and 6.1, which individually have computational complexity of ${\cal O}\left( N \right)$, ${\cal O}\left( K \right)$, and ${\cal O}\left( K \right)$ in each iteration. The computational complexity of the remaining two subproblems remains constant throughout the algorithm. Let $r$ represent the number of iterations required for {\bf{Algorithm 1}} to converge. Consequently, the overall computational complexity of {\bf{Algorithm 1}} can be expressed as ${\cal O}\left( {r\left( {N + 2K} \right)} \right)$. The computational complexity of the consensus ADMM-based beamforming algorithm is significantly lower compared to that of the interior point method used to solve SDP problems. Therefore, the proposed algorithm is well-suited for a wide range of complex scenarios, enabling efficient and practical implementations.
\section{Simulation Results}
In this section, we numerically evaluate the performance of our proposed beamforming design algorithm in the considered transmissive RIS transceiver enabled downlink communication system. In the simulatiuon setup, we consider a scenario where an RIS transceiver equipped with 16 transmissive phase control elements is located at (0m, 0m, 15m), and 5 users are randomly distributed in a circle whose center coordinates as (0m, 0m, 0m) with a radius of 50m. The antenna spacing is set to half the wavelength of the carrier. The maximum available transmissive power of each element is ${{P}_{t}=1}$mW. The widely adopted distance-dependent path-loss model is also employed in our system, i.e., the RIS-user link is assumed to follow Rician fading with a Rician factor and a path loss exponent. The noise power at all users is set as ${{\sigma }^{2}}=-50$dBm. The path loss exponent, denoted as $\alpha$, is set to 3. The path loss $\beta$ is configured as -20dB at a reference distance of 1m. Additionally, the Rician factor $\kappa$ is set to 3dB. The threshold for algorithm convergence is set as $\varepsilon ={{10}^{-3}}$.

We first illustrate the convergence performance of our proposed algorithm via Fig. 3. Fig. 3 demonstrates the rapid convergence of the max-min SINR within approximately 4 iterations, providing evidence of the favorable convergence property of our proposed algorithm. Moreover, we conduct a comparative analysis to assess the influence of different numbers of transmissive elements on the system performance. In this work, the considered RIS deployment adopts the UPA configuration, where the number of elements in the horizontal and vertical directions is identical. Consequently, we evaluated the performance of the proposed beamforming design algorithm in terms of the max-min SINR, considering scenarios with 9, 16, 25 and 36 RIS transmissive elements. The results indicate that a larger number of RIS transmissive elements leads to higher max-min SINR. This behavior can be attributed to the increased diversity gain achieved by leveraging a larger number of RIS transmissive elements, resembling the performance trend observed in conventional multi-antenna transceiver architectures.
\begin{figure}
	\centerline{\includegraphics[width=9cm]{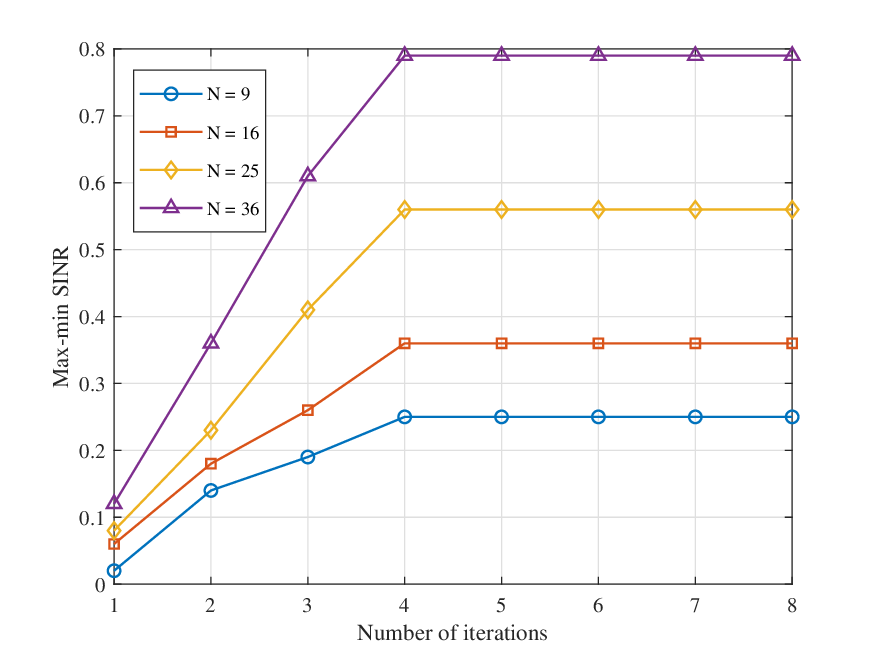}}
	\caption{Convergence of beamforming algorithm based on consensus ADMM.}
	\label{Fig3}
\end{figure}

Subsequently, we delve into the analysis of the relationship between the maximum available transmit power of the RIS transmission element and the corresponding max-min SINR. In the Fig. 4, it can be observed that the max-min SINR exhibits a consistent upward trend as the maximum available transmit power increases. This behavior can be attributed to the fact that an augmented maximum transmission power empowers the system to deliver enhanced diversity gain to the users, thereby resulting in an amplified max-min SINR. When the maximum available transmit power of the transmissive element is the same, the trend of max-min SINR is consistent with the above mentioned.
\begin{figure}
	\centerline{\includegraphics[width=9cm]{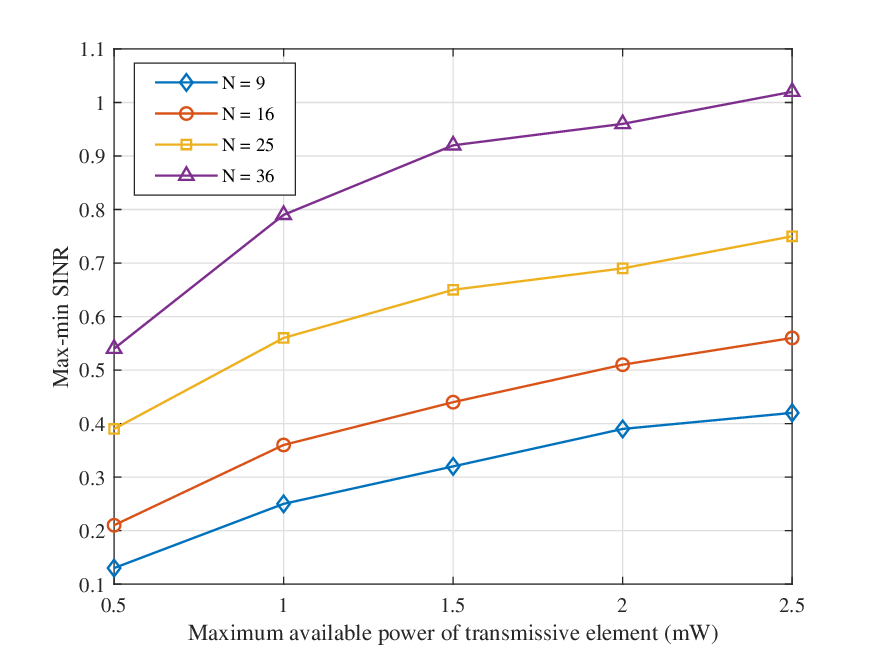}}
	\caption{System achievable sum-rate versus maximum power of each transmissive element.}
	\label{Fig4}
\end{figure}

Next, Fig. 5 showcases the interplay between max-min SINR and the number of RIS transmissive elements under different Rician factors. Remarkably, it is observed that the max-min SINR experienced by all users increases as the number of RIS transmissive elements escalates. The amplified diversity gain observed in the system with an increased number of RIS transmissive elements can be attributed to the enhanced diversity of the newly proposed transceiver architecture presented in this paper. Additionally, when the number of transmissive elements remains constant, the max-min SINR exhibits an upward trend with an increasing Rician factor. This behavior is primarily due to the improved LoS performance of the channel associated with a larger Rician factor, resulting in a greater overall performance gain.
\begin{figure}
	\centerline{\includegraphics[width=9cm]{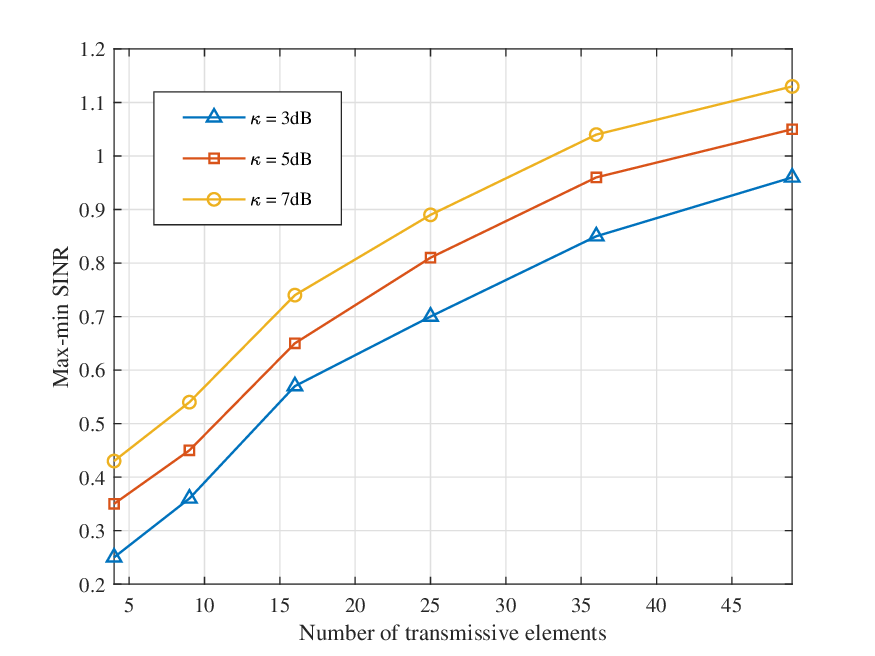}}
	\caption{Max-min SINR versus the number of transmissive elements.}
	\label{Fig5}
\end{figure}



Then, Fig. 6 illustrates the relationship between the max-min SINR and different Rician factors at the maximum transmit power available for RIS elements. It is evident from the plot that as the Rician factor increases, the max-min SINR exhibits a continuous rise. This can be attributed to the heightened presence of the LoS component in the channel, which enhances the service quality for each user and consequently leads to an increase in the max-min SINR. Moreover, when the Rician factor remains constant, the max-min SINR experiences an upward trend with the increment of the maximum available transmit power of the RIS elements.
\begin{figure}
	\centerline{\includegraphics[width=9cm]{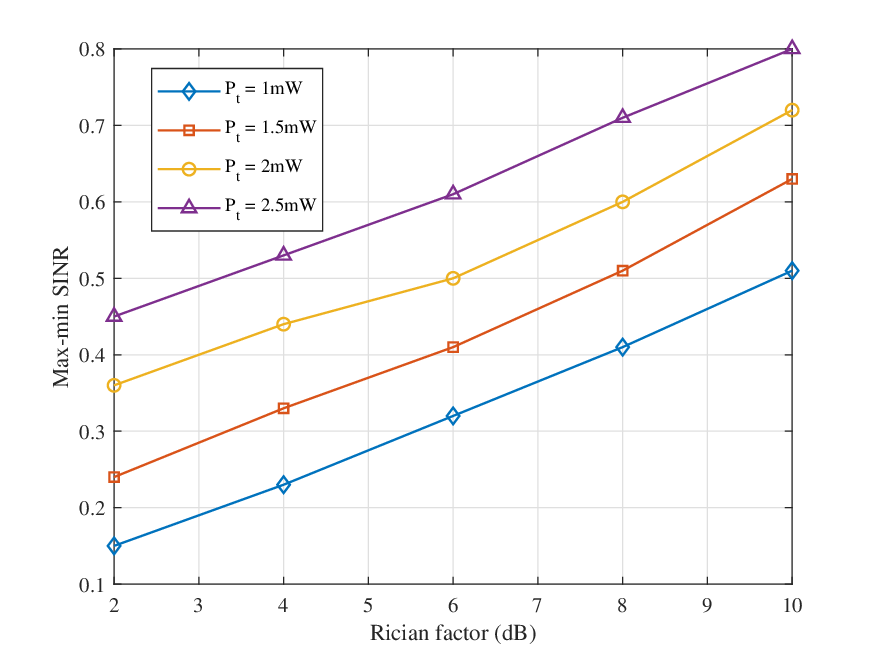}}
	\caption{Max-min SINR versus Rician factor.}
	\label{Fig6}
\end{figure}

Finally, in Fig. 7 and Fig. 8, we present a comparison of the computation time between the proposed beamforming algorithm based on consensus ADMM and the SDP algorithm solved by the interior point method, under different numbers of RIS transmissive elements and different numbers of users, respectively. As depicted in the Fig.7 and Fig. 8, it is evident that the proposed algorithm exhibits superior time efficiency compared to the interior point method. This observation aligns with the earlier analysis on computational complexity. Furthermore, the Fig.7 and Fig. 8 also demonstrate that as the number of RIS transmissive elements and the number of users increase, the efficiency advantage of the proposed consensus ADMM beamforming algorithm becomes more prominent. This characteristic renders it well-suited for complex network environments in the future, thereby enhancing its practicality.
\begin{figure}
	\centerline{\includegraphics[width=9cm]{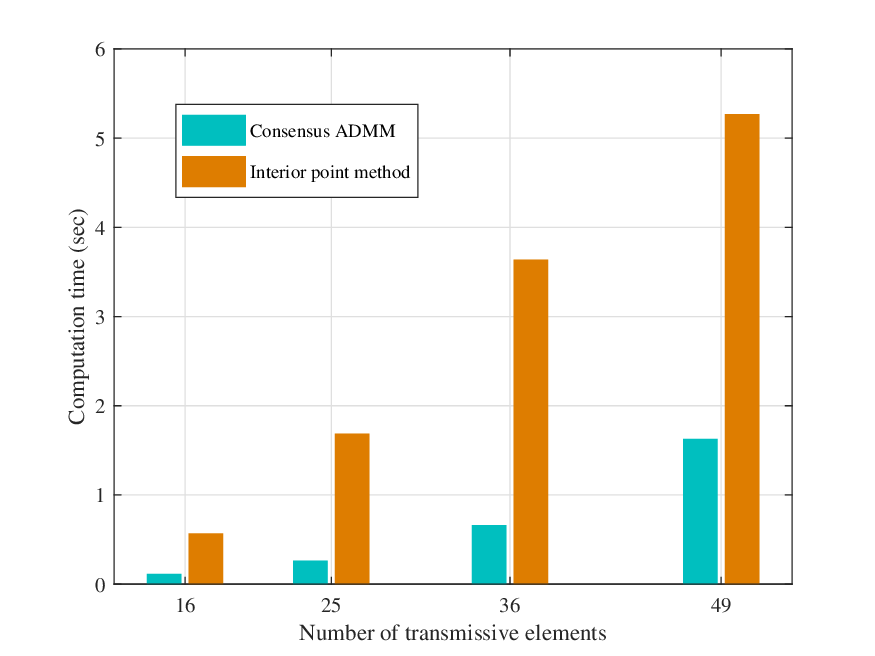}}
	\caption{Computation time versus the number of transmissive elements.}
	\label{Fig7}
\end{figure}

\begin{figure}
	\centerline{\includegraphics[width=9cm]{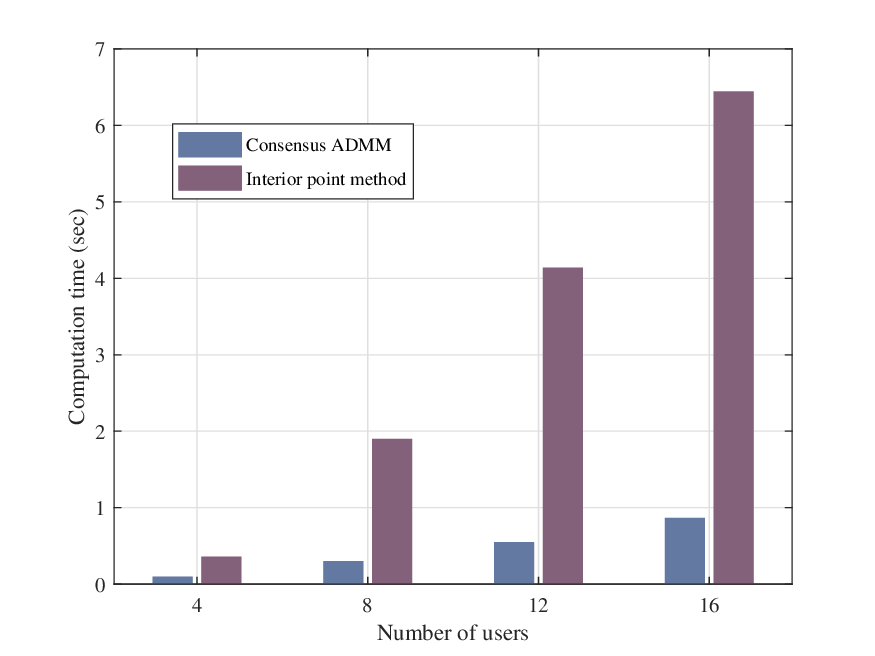}}
	\caption{Computation time versus the number of users.}
	\label{Fig8}
\end{figure}

\begin{figure*}[!t]
	\normalsize
	\setcounter{my3}{\value{equation}}
	\setcounter{equation}{67}
        \begin{equation}
	\begin{aligned}
		\left\| {{{\bf{\Gamma }}_n} - {\bf{F}} + {{\bf{\Xi }}_n}} \right\|_F^2 = {\left( {{{\left( {{{\bf{\Gamma }}_n}} \right)}_{1,1}} - {{\bf{F}}_{1,1}} + {{\left( {{{\bf{\Xi }}_n}} \right)}_{1,1}}} \right)^ * }\left( {{{\left( {{{\bf{\Gamma }}_n}} \right)}_{1,1}} - {{\bf{F}}_{1,1}} + {{\left( {{{\bf{\Xi }}_n}} \right)}_{1,1}}} \right) +  \cdots \\
		 + {\left( {{{\left( {{{\bf{\Gamma }}_n}} \right)}_{1,K}} - {{\bf{F}}_{1,K}} + {{\left( {{{\bf{\Xi }}_n}} \right)}_{1,K}}} \right)^ * }\left( {{{\left( {{{\bf{\Gamma }}_n}} \right)}_{1,K}} - {{\bf{F}}_{1,K}} + {{\left( {{{\bf{\Xi }}_n}} \right)}_{1,K}}} \right) + \cdots\\
	    {\rm{ + }}{\left( {{{\left( {{{\bf{\Gamma }}_n}} \right)}_{N,1}} - {{\bf{F}}_{N,1}} + {{\left( {{{\bf{\Xi }}_n}} \right)}_{N,1}}} \right)^ * }\left( {{{\left( {{{\bf{\Gamma }}_n}} \right)}_{N,1}} - {{\bf{F}}_{1,1}} + {{\left( {{{\bf{\Xi }}_n}} \right)}_{N,1}}} \right) +  \cdots \\
	    + {\left( {{{\left( {{{\bf{\Gamma }}_n}} \right)}_{N,K}} - {{\bf{F}}_{N,K}} + {{\left( {{{\bf{\Xi }}_n}} \right)}_{N,K}}} \right)^ * }\left( {{{\left( {{{\bf{\Gamma }}_n}} \right)}_{N,K}} - {{\bf{F}}_{N,K}} + {{\left( {{{\bf{\Xi }}_n}} \right)}_{N,K}}} \right).
	\end{aligned}
        \end{equation}
        \setcounter{equation}{73}
        \begin{equation}
	\begin{aligned}
		{\left| {{\bf{\tilde h}}_k^H\left( {{\rm{vec}}\left( {{{\bf{\Psi }}_k}} \right) \circ {{\bf{a}}_k}} \right)} \right|^2}& = {\left( {{\rm{vec}}\left( {{{\bf{\Psi }}_k}} \right) \circ {{\bf{a}}_k}} \right)^H}{{{\bf{\tilde h}}}_k}{\bf{\tilde h}}_k^H\left( {{\rm{vec}}\left( {{{\bf{\Psi }}_k}} \right) \circ {{\bf{a}}_k}} \right)\\
		&= \left( {{\rm{vec}}\left( {{{\bf{\Psi }}_k}} \right)} \right)_1^ * \left( {{{\bf{a}}_k}} \right)_1^ * {\left( {{{\bf{H}}_k}} \right)_{1,1}}{\left( {{\rm{vec}}\left( {{{\bf{\Psi }}_k}} \right)} \right)_1}{\left( {{{\bf{a}}_k}} \right)_1} +  \cdots \\
		& + \left( {{\rm{vec}}\left( {{{\bf{\Psi }}_k}} \right)} \right)_1^ * \left( {{{\bf{a}}_k}} \right)_1^ * {\left( {{{\bf{H}}_k}} \right)_{1,NK}}{\left( {{\rm{vec}}\left( {{{\bf{\Psi }}_k}} \right)} \right)_{NK}}{\left( {{{\bf{a}}_k}} \right)_{NK}}\\
		&+  \cdots \\
		& + \left( {{\rm{vec}}\left( {{{\bf{\Psi }}_k}} \right)} \right)_{NK}^ * \left( {{{\bf{a}}_k}} \right)_{NK}^ * {\left( {{{\bf{H}}_k}} \right)_{NK,1}}{\left( {{\rm{vec}}\left( {{{\bf{\Psi }}_k}} \right)} \right)_1}{\left( {{{\bf{a}}_k}} \right)_1} +  \cdots \\
		& + \left( {{\rm{vec}}\left( {{{\bf{\Psi }}_k}} \right)} \right)_{NK}^ * \left( {{{\bf{a}}_k}} \right)_{NK}^ * {\left( {{{\bf{H}}_k}} \right)_{NK,NK}}{\left( {{\rm{vec}}\left( {{{\bf{\Psi }}_k}} \right)} \right)_{NK}}{\left( {{{\bf{a}}_k}} \right)_{NK}},
	\end{aligned}
        \end{equation} 
        \begin{equation}
	\begin{aligned}
		&\frac{{\partial {{\left| {{\bf{\tilde h}}_k^H\left( {{\rm{vec}}\left( {{{\bf{\Psi }}_k}} \right) \circ {{\bf{a}}_k}} \right)} \right|}^2}}}{{\partial {{\left( {{\rm{vec}}\left( {{{\bf{\Psi }}_k}} \right)} \right)}_1}}} = \left( {{\rm{vec}}\left( {{{\bf{\Psi }}_k}} \right)} \right)_1^ * \left( {{{\bf{a}}_k}} \right)_1^ * {\left( {{{\bf{a}}_k}} \right)_1}{\left( {{{\bf{H}}_k}} \right)_{1,1}} +  \cdots + \left( {{\rm{vec}}\left( {{{\bf{\Psi }}_k}} \right)} \right)_{NK}^ * \left( {{{\bf{a}}_k}} \right)_{NK}^ * {\left( {{{\bf{a}}_k}} \right)_1}{\left( {{{\bf{H}}_k}} \right)_{NK,1}},\\
		&\frac{{\partial {{\left| {{\bf{\tilde h}}_k^H\left( {{\rm{vec}}\left( {{{\bf{\Psi }}_k}} \right) \circ {{\bf{a}}_k}} \right)} \right|}^2}}}{{\partial {{\left( {{\rm{vec}}\left( {{{\bf{\Psi }}_k}} \right)} \right)}_2}}} = \left( {{\rm{vec}}\left( {{{\bf{\Psi }}_k}} \right)} \right)_1^ * \left( {{{\bf{a}}_k}} \right)_1^ * {\left( {{{\bf{a}}_k}} \right)_2}{\left( {{{\bf{H}}_k}} \right)_{1,2}} +  \cdots + \left( {{\rm{vec}}\left( {{{\bf{\Psi }}_k}} \right)} \right)_{NK}^ * \left( {{{\bf{a}}_k}} \right)_{NK}^ * {\left( {{{\bf{a}}_k}} \right)_2}{\left( {{{\bf{H}}_k}} \right)_{NK,2}},\\
		& \vdots \\
		&\frac{{\partial {{\left| {{\bf{\tilde h}}_k^H\left( {{\rm{vec}}\left( {{{\bf{\Psi }}_k}} \right) \circ {{\bf{a}}_k}} \right)} \right|}^2}}}{{\partial {{\left( {{\rm{vec}}\left( {{{\bf{\Psi }}_k}} \right)} \right)}_{NK}}}} = \left( {{\rm{vec}}\left( {{{\bf{\Psi }}_k}} \right)} \right)_1^ * \left( {{{\bf{a}}_k}} \right)_1^ * {\left( {{{\bf{a}}_k}} \right)_{NK}}{\left( {{{\bf{H}}_k}} \right)_{1,NK}} +  \cdots + \left( {{\rm{vec}}\left( {{{\bf{\Psi }}_k}} \right)} \right)_{NK}^ * \left( {{{\bf{a}}_k}} \right)_{NK}^ * {\left( {{{\bf{a}}_k}} \right)_{NK}}{\left( {{{\bf{H}}_k}} \right)_{NK,NK}}.
	\end{aligned}
\end{equation}
\setcounter{equation}{\value{my3}}
\hrulefill
\vspace*{4pt}
\end{figure*}
\section{Conclusions}
In this paper, we investigate the novel architecture of a transmissive RIS transceiver to facilitate download multi-stream communication. The RIS is seamlessly integrated with an intelligent controller, thereby establishing a comprehensive framework. In order to ensure the fairness of all users, our objective lies in the meticulous design of a beamforming matrix that maximizes the minimum SINR while considering the constraint imposed by the maximum available transmission power of the RIS elements. Given the inherent complexity arising from the interdependence of variables, we present a linear-complexity beamforming scheme based on consensus ADMM algorithm. The algorithm effectively decouples the variables and resolves the resultant intricate non-convex optimization problem. Through extensive simulations, we substantiate the efficacy of the proposed algorithm within the context of our proposed scheme. The design scheme proposed in this work, along with the obtained results, showcases the potential of transmissive RIS as a novel architecture for communication systems. It offers a pathway to realize multi-antenna transceivers that are characterized by their low cost and low power consumption. This development is highly attractive and holds significant promise for future wireless communications. Our future research endeavors encompass the extension of our work to encompass a broader range of practical communication scenarios, thus further enhancing its applicability.
%
%
%
\appendices
\section{Proof of Eq. (33)}
The objective function of problem P4.1 can be written as Eq. (68). The partial derivative of $\left\| {{{\bf{\Gamma }}_n} - {\bf{F}} + {{\bf{\Xi }}_n}} \right\|_F^2$ w.r.t each component of matrix ${{{\bf{\Gamma }}_n}}$ can be expressed as
\setcounter{equation}{68}
\begin{equation}
	\begin{aligned}
		&\frac{{\partial \left\| {{{\bf{\Gamma }}_n} - {\bf{F}} + {{\bf{\Xi }}_n}} \right\|_F^2}}{{\partial {{\left( {{{\bf{\Gamma }}_n}} \right)}_{1,1}}}} = \left( {{{\bf{\Gamma }}_n}} \right)_{1,1}^ *  - {\bf{F}}_{1,1}^ *  + \left( {{{\bf{\Xi }}_n}} \right)_{1,1}^ * ,\\
		&\frac{{\partial \left\| {{{\bf{\Gamma }}_n} - {\bf{F}} + {{\bf{\Xi }}_n}} \right\|_F^2}}{{\partial {{\left( {{{\bf{\Gamma }}_n}} \right)}_{1,2}}}} = \left( {{{\bf{\Gamma }}_n}} \right)_{1,2}^ *  - {\bf{F}}_{1,2}^ *  + \left( {{{\bf{\Xi }}_n}} \right)_{1,2}^ * ,\\
		& \vdots \\
		&\frac{{\partial \left\| {{{\bf{\Gamma }}_n} - {\bf{F}} + {{\bf{\Xi }}_n}} \right\|_F^2}}{{\partial {{\left( {{{\bf{\Gamma }}_n}} \right)}_{1,K}}}} = \left( {{{\bf{\Gamma }}_n}} \right)_{1,K}^ *  - {\bf{F}}_{1,K}^ *  + \left( {{{\bf{\Xi }}_n}} \right)_{1,K}^ * ,\\
		& \vdots \\
		&\frac{{\partial \left\| {{{\bf{\Gamma }}_n} - {\bf{F}} + {{\bf{\Xi }}_n}} \right\|_F^2}}{{\partial {{\left( {{{\bf{\Gamma }}_n}} \right)}_{N,K}}}} = \left( {{{\bf{\Gamma }}_n}} \right)_{N,K}^ *  - {\bf{F}}_{N,K}^ *  + \left( {{{\bf{\Xi }}_n}} \right)_{N,K}^ * .
	\end{aligned}
\end{equation}
Thus, we have
\begin{equation}
	\frac{{\partial \left\| {{{\bf{\Gamma }}_n} - {\bf{F}} + {{\bf{\Xi }}_n}} \right\|_F^2}}{{\partial {{\bf{\Gamma }}_n}}} = {\bf{\Gamma }}_n^ *  - {{\bf{F}}^ * } + {\bf{\Xi }}_n^ * .
\end{equation}

Additionally, ${\left\| {{\rm{vec}}\left( {{{\bf{\Gamma }}_n}} \right) \circ {{\bf{b}}_n}} \right\|^2}$ can be written as
\begin{equation}
\begin{aligned}
     {\left\| {{\rm{vec}}\left( {{{\bf{\Gamma }}_n}} \right) \circ {{\bf{b}}_n}} \right\|^2} = {\left| {{{\left( {{\rm{vec}}\left( {{{\bf{\Gamma }}_n}} \right)} \right)}_1}{{\left( {{{\bf{b}}_n}} \right)}_1}} \right|^2} + \\{\left| {{{\left( {{\rm{vec}}\left( {{{\bf{\Gamma }}_n}} \right)} \right)}_2}{{\left( {{{\bf{b}}_n}} \right)}_2}} \right|^2} +  \cdots  + {\left| {{{\left( {{\rm{vec}}\left( {{{\bf{\Gamma }}_n}} \right)} \right)}_{NK}}{{\left( {{{\bf{b}}_n}} \right)}_{NK}}} \right|^2}.
\end{aligned}
\end{equation}
The partial derivative of ${\left\| {{\rm{vec}}\left( {{{\bf{\Gamma }}_n}} \right) \circ {{\bf{b}}_n}} \right\|^2}$ w.r.t each component of matrix $\bf{F}$ can be given by
\begin{equation}
	\begin{aligned}
		&\frac{{\partial {{\left\| {{\rm{vec}}\left( {{{\bf{\Gamma }}_n}} \right) \circ {{\bf{b}}_n}} \right\|}^2}}}{{\partial {{\left( {{{\bf{\Gamma }}_n}} \right)}_{1,1}}}} = {\left| {{{\left( {{{\bf{b}}_n}} \right)}_1}} \right|^2}\left( {{\rm{vec}}\left( {{{\bf{\Gamma }}_n}} \right)} \right)_1^ * ,\\
		&\frac{{\partial {{\left\| {{\rm{vec}}\left( {{{\bf{\Gamma }}_n}} \right) \circ {{\bf{b}}_n}} \right\|}^2}}}{{\partial {{\left( {{\bf{\Gamma _n}}} \right)}_{2,1}}}} = {\left| {{{\left( {{{\bf{b}}_n}} \right)}_2}} \right|^2}\left( {{\rm{vec}}\left( {{{\bf{\Gamma }}_n}} \right)} \right)_2^ * ,\\
		& \vdots \\
		&\frac{{\partial {{\left\| {{\rm{vec}}\left( {{{\bf{\Gamma }}_n}} \right) \circ {{\bf{b}}_n}} \right\|}^2}}}{{\partial {{\left( {{\bf{\Gamma _n}}} \right)}_{N,K}}}} = {\left| {{{\left( {{{\bf{b}}_n}} \right)}_{NK}}} \right|^2}\left( {{\rm{vec}}\left( {{{\bf{\Gamma }}_n}} \right)} \right)_{NK}^ * .
	\end{aligned}
\end{equation}
Therefore, we have
\begin{equation}
	\frac{{\partial {{\left\| {{\rm{vec}}\left( {{{\bf{\Gamma }}_n}} \right) \circ {{\bf{b}}_n}} \right\|}^2}}}{{\partial {{\bf{\Gamma }}_n}}} = {{\bf{B}}_n} \circ {\bf{\Gamma }}_n^ * .
\end{equation}

According to Eq. (70) and Eq. (73), the Eq. (33) can be obtained.

\section{Proof of Eq. (44) and Eq. (46)}
For the Eq. (44), ${\left| {{\bf{\tilde h}}_k^H\left( {{\rm{vec}}\left( {{{\bf{\Psi }}_k}} \right) \circ {{\bf{a}}_k}} \right)} \right|^2}$ can be written as Eq. (74), where ${{\bf{H}}_k} = {{{\bf{\tilde h}}}_k}{\bf{\tilde h}}_k^H \in {\mathbb{C}^{NK \times NK}}$. The partial derivative of ${\left| {{\bf{\tilde h}}_k^H\left( {{\rm{vec}}\left( {{{\bf{\Psi }}_k}} \right) \circ {{\bf{a}}_k}} \right)} \right|^2}$ w.r.t each component of vector ${{\rm{vec}}\left( {{{\bf{\Psi }}_k}} \right)}$ can be expressed as Eq. (75). Thus, 
\setcounter{equation}{75}
\begin{equation}
	\frac{{\partial {{\left| {{\bf{\tilde h}}_k^H\left( {{\rm{vec}}\left( {{{\bf{\Psi }}_k}} \right) \circ {{\bf{a}}_k}} \right)} \right|}^2}}}{{\partial {\rm{vec}}\left( {{{\bf{\Psi }}_k}} \right)}} = \left( {{\bf{H}}_k^T \circ {{\bf{A}}_k}} \right){\rm{vec}}\left( {{\bf{\Psi }}_k^ * } \right),\forall k.
\end{equation}
By applying SCA, the lower bound of ${\left| {{\bf{\tilde h}}_k^H\left( {{\rm{vec}}\left( {{{\bf{\Psi }}_k}} \right) \circ {{\bf{a}}_k}} \right)} \right|^2}$ can be obtained in Eq. (44). Similarly, Eq. (46) can also be obtained accordingly.



\ifCLASSOPTIONcaptionsoff
  \newpage
\fi




\bibliographystyle{IEEEtran}
\bibliography{reference}

\end{document}